\documentclass[reqno]{amsproc}
\usepackage[latin1]{inputenc}
\usepackage{latexsym}
\usepackage[colorlinks, citecolor=blue,linkcolor=black]{hyperref}
\hypersetup{
	plainpages=false,
	colorlinks=true,
	linkcolor=blue, 
	anchorcolor=black, 
	citecolor=blue, 
	urlcolor=blue, 
	menucolor=black, 
	filecolor=black, 
	bookmarksopen=true,
	bookmarksnumbered=true}
\usepackage{cite}
\usepackage{xcolor}
\usepackage{mathrsfs,amstext,amsmath,amssymb,amsfonts,bm}
\def\dprod{\displaystyle\prod}

\def\dsum{\displaystyle\sum}

\newtheorem{definition}{Definition}
\newtheorem{example}[definition]{Example}

\newtheorem{remark}[definition]{Remark}

\makeatletter
\@addtoreset{definition}{section}
\@addtoreset{equation}{section}
\makeatother

\allowdisplaybreaks[4]

\title{Characterization of the  $W_{1+\infty}$-n-algebra and applications }
\author{Fridolin Melong \and  Raimar Wulkenhaar}
\address[]{
	International Chair in Mathematical Physics and Applications,
	ICMPA-UNESCO Chair,\newline University of Abomey-Calavi, 072 BP 50, Cotonou, Rep. of Benin,
{\itshape e-mail:} \normalfont\texttt{fridomelong@gmail.com}}
\address[]{Mathematisches Institut der
	Universit\"at M\"unster \newline
	Einsteinstr.\ 62, 48149 M\"unster, Germany, 
	{\itshape e-mail:} \normalfont
	\texttt{raimar@math.uni-muenster.de}}
\begin{document}	
	\begin{abstract}
	In this paper, we construct the  $W_{1+\infty}$-n-algebras in the framework of the  generalized quantum algebra. We characterize the  $\mathcal{R}(p,q)$-multi-variable $W_{1+\infty}$-algebra  and derive its $n$-algebra which is the generalized Lie algebra for $n$ even. Furthermore,  we investigate the $\mathcal{R}(p,q)$-elliptic hermitian matrix model and determine a toy model  for the generalized quantum $W_{\infty}$ constraints. Also, we deduce  particular cases of our results.
	\end{abstract}	
\maketitle
\tableofcontents
	\section{Introduction}
	The $W_{1+\infty}$ algebra is an important infinite dimensional
	algebra  which is the higher-spin extension
	of the Virasoro algebra \cite{BL1,BL2,BL3}. It has attracted
	much interest from physical and mathematical
	points of view. In the context of integrable systems,
	the $M_{1+\infty}$ algebra was found to be associated
	with the first Hamiltonian structure of the KP
	hierarchy \cite{GU}, \cite{Estienne}. It is well-known that Calogero-Moser-Sutherland (CMS) models are  one-dimensional
	many-body integrable systems, which have received
	lots of attention\cite{Hasebe14},\cite{Hasebe}\cite{Neupert}. The $SU(\nu)$ Calogero spin system
	is the generalization of the quantum Calegero
	model. In this integrable system, Hikami and Wadati
	found that there exists a higher symmetry, i.e.,
	$W_{1+\infty}$ algebra \cite{CFZ}. This $W$ algebraic structure unifies
	the Calogero type and the Sutherland type (trigonometric
	potential).
	Recently great attention has been paid to $n$-algebra, especially $3$-algebra due to a world-volume
	description of multiple $M2$-branes proposed by Bagger
	and Lambert\cite{BL1} and Gustavsson \cite{GU}. One also
	found  applications of $n$-algebra in integrable
	systems and condensed matter physics.
	 
	Furthermore, the notion of matrix models have attracted attention since they provide a large set of  physical systems. In the literature, the Virasoro constraints for the partition function play a fundamental role to  understand the structure of the hermitian matrix models\cite{DVV, MM}. Thus, the $W$ constraints in matrix models have also attract more attention \cite{G,IM,MS}.  Recently the trigonometric and elliptic generalizations of hermitian matrix models have also found  interest in the field of supersymmetric gauge theories \cite{AKMV,DO,GY,M}. 
	
	Besides, the notion of quantum algebras was 
	 very important in the areas of mathematical physics and others. Then, recently the generalized quantum algebras were introduced in \cite{HB1}. From this, the generalizations of the Virasoro algebra, conformal Virasoro algebra,  Heisenberg Virasoro algebra, and matrix models were investigated. For more details, the reader is refered to \cite{HM,HMM}.
	
	Let $p$ and $q$ be two positive real numbers such that $ 0<q<p<1.$ We consider a meromorphic function ${\mathcal R}$ defined on $\mathbb{C}\times\mathbb{C}$ by\cite{HB}:\begin{equation}\label{r10}
	\mathcal{R}(u,v)= \sum_{s,t=-l}^{\infty}r_{st}u^sv^t,
	\end{equation}
	with a possible isolated singularity at zero, 
	where $r_{st}$ are complex numbers, $l\in\mathbb{N}\cup\left\lbrace 0\right\rbrace,$ $\mathcal{R}(p^n,q^n)>0,  \forall n\in\mathbb{N},$ and $\mathcal{R}(1,1)=0$ by definition. We denote by $\mathbb{D}_{R}$ the bidisk \begin{eqnarray*}
		\mathbb{D}_{R}
		=\left\lbrace w=(w_1,w_2)\in\mathbb{C}^2: |w_j|<R_{j} \right\rbrace,
	\end{eqnarray*}
	where $R$ is the convergence radius of the series (\ref{r10}) defined by Hadamard formula as follows:
	\begin{eqnarray*}
		\limsup_{s+t \longrightarrow \infty} \sqrt[s+t]{|r_{st}|R^s_1\,R^t_2}=1.
	\end{eqnarray*}
	For the proof and more details see \cite{TN}. 
	We define the  $\mathcal{R}(p,q)$-numbers  \cite{HB}:
	\begin{equation*}
	[n]_{\mathcal{R}(p,q)}:=\mathcal{R}(p^n,q^n),\quad n\in\mathbb{N},
	\end{equation*}
	the
	$\mathcal{R}(p,q)$-factorials by:
	\begin{eqnarray*}\label{s0}
		[n]!_{\mathcal{R}(p,q)}:=\left \{
		\begin{array}{l}
			1\quad\mbox{for}\quad n=0\\
			\\
			\mathcal{R}(p,q)\cdots\mathcal{R}(p^n,q^n)\quad\mbox{for}\quad n\geq 1,
		\end{array}
		\right .
	\end{eqnarray*}
	and the  $\mathcal{R}(p,q)$-binomial coefficients
	\begin{eqnarray*}\label{bc}
		\mathbb{C}^{k}_{n} := \frac{[n]!_{\mathcal{R}(p,q)}}{[k]!_{\mathcal{R}(p,q)}[n-k]!_{\mathcal{R}(p,q)}},\quad n\geq k.
	\end{eqnarray*}
	We denote by 
	${\mathcal O}(\mathbb{D}_R)$ the set of holomorphic functions defined
	on $\mathbb{D}_R$ and 
	consider the following linear operators defined on  $\mathcal{O}(\mathbb{D}_{R}),$ (see \cite{HB1} for more details),
	\begin{eqnarray*}
		\; P:\Psi\longmapsto P\Psi(z):&=& \Psi(pz),\\
		\;Q:\Psi\longmapsto Q\Psi(z):&=& \Psi(qz),
	\end{eqnarray*}
	and the $\mathcal{R}(p,q)$-derivative \cite{HB}
	\begin{eqnarray}\label{r5}
	{\mathcal D}_{\mathcal{R}( p,q)}:={\mathcal D}_{p,q}\frac{p-q}{P-Q}\mathcal{R}( P,Q)=\frac{p-q}{p^{P}-q^{Q}}\mathcal{R}(p^{P},q^{Q}){\mathcal D}_{p,q}
	\end{eqnarray}
	where ${\mathcal D}_{p,q}$ is the $(p,q)$-derivative:
	\begin{eqnarray*}
		{\mathcal D}_{p,q}\Psi(z):=\frac{\Psi(pz)-\Psi(qz)}{z(p-q)}.
	\end{eqnarray*}
	
	The  algebra associated with the $\mathcal{R}(p,q)$-deformation is a quantum algebra, denoted $\mathcal{A}_{\mathcal{R}(p,q)},$ generated by the set of operators $\{1, A, A^{\dagger}, N\}$ satisfying the following commutation relations:
	\begin{eqnarray*}
	&& \label{algN1}
	\quad A A^\dag= [N+1]_{\mathcal {R}(p,q)},\quad\quad\quad A^\dag  A = [N]_{\mathcal {R}(p,q)}.
	\cr&&\left[N,\; A\right] = - A, \qquad\qquad\quad \left[N,\;A^\dag\right] = A^\dag
	\end{eqnarray*}
	with the realization on  ${\mathcal O}(\mathbb{D}_R)$ given by:
	\begin{eqnarray*}\label{algNa}
		A^{\dagger} := z,\qquad A:=\partial_{\mathcal {R}(p,q)}, \qquad N:= z\partial_z,
	\end{eqnarray*} 
	where $\partial_z:=\frac{\partial}{\partial z}$ is the  derivative on $\mathbb{C}.$
	  
	  The aims of this work are to generalize the $W_{1+\infty}$ algebra and its $n$-algebra generated by the quantum algebra \cite{HB1}. The notions of elliptic hermitian matrix model and a toy model for the $\mathcal{R}(p,q)$-$W_{\infty}$ constraints are determined and presented as applications.
	  
	This paper is organized as follows: section $2$ is reserved to the realization of the  $W_{1+\infty}$ algebra,  its n-algebras and deduce particular cases. 
 Section $3$ is dedicated to another characterization of
the  $W_{1+\infty}$-n-algebra. In  section $4$,  we investigate the $\mathcal{R}(p,q)$-multi-variable $W_{1+\infty}$-n- algebra. In section $5$, we present the $\mathcal{R}(p,q)$-$W_{\infty}$ constraints for the elliptic hermitian matrix model. Section $6$ concerns the $\mathcal{R}(p,q)$-$W_{\infty}$ constraints for a toy model.  Finally, particular cases are deduced and concluding remarks in section $7$.
\section{Realization of the $\mathcal{R}(p,q)$-$W_{1+\infty}$-$n$ algebra}
In this section, we construct the $W_{1+\infty}$ $n$-algebra from the $\mathcal{R}(p,q)$-deformed quantum algebra introduced in \cite{HB1}. Particular cases are also deduced.

The generalized quantum Leibniz formula is given by:
\begin{equation*}\label{RpqRL}
\mathcal{D}^r_{\mathcal{R}(p,q)}\big(f(x)g(x)\big)=\sum_{j=0}^{r}\mathbb{C}^{j}_{r}\mathcal{D}^{j}_{\mathcal{R}(p,q)}\big(f(xq^{r-j})\big)\mathcal{D}^{r-j}_{\mathcal{R}(p,q)}g\big(xp^{j}\big).
\end{equation*}
We consider the $\mathcal{R}(p,q)$-operators defined by:
\begin{eqnarray}\label{Rpqop1}
W_m^r=z^{m+r-1}\mathcal{D}_{\mathcal{R}(p,q)}^{r-1}, \ \ r\in \mathbb{Z}_+,\, m \in\mathbb{N}.
\end{eqnarray}
 Then, 
the $\mathcal{R}(p,q)$-$W_{1+\infty}$ algebra is generated by the operators \eqref{Rpqop1}  satisfying the  relation:
\begin{align}\label{Rpqwalg}
\left[W_{m}^{r}, W_{n}^{s}\right]
&=K(P,Q)\bigg(\dsum_{j=0}^{r-1}q^{(r-1-j)(n+s-1)}p^{j}\mathbb{C}_{r-1}^{j}\mathbb{A}_{n+s-1}^{j}\nonumber\\
&-\dsum_{j=0}^{s-1}q^{(s-1-j)(m+r-1)}p^{j}\mathbb{C}_{s-1}^{j}
\mathbb{A}_{m+r-1}^{j}\bigg)W_{m+n}^{r+s-1-j},
\end{align}
where $K(P,Q)=\frac{p-q}{p^{P}-q^{Q}}\mathcal{R}(p^{P},q^{Q})$ and 
$$\mathbb{A}_n^{k}=\left\{\begin{array}{cc}
[n]_{\mathcal{R}(p,q)}[n-1]_{\mathcal{R}(p,q)}\cdots[n-k+1]_{\mathcal{R}(p,q)},& k\leqslant n,\\
0,                  &k>n.\end{array}\right. $$
\begin{remark}
	Note that, by taking $r=s=2$ in the relation \eqref{Rpqwalg}, we derive the $\mathcal{R}(p,q)$-algebra:
	\begin{equation*}
	\left[W_{m}^2, W_{n}^2\right]=K(P,Q)\bigg(\big(q^{n+1}-q^{m+1}\big)W_{m+n}^3+\big([n+1]_{\mathcal{R}(p,q)}-[m+1]_{\mathcal{R}(p,q)}\big)\bigg)W_{m+n}^2.
	\end{equation*}
\end{remark} 
We consider the $n$-commutator of the  $\mathcal{R}(p,q)$-operators \eqref{Rpqop1}  as follows:
\begin{equation}\label{Rpqn-bracket}
\left[W_{m_1}^{r_1}, W_{m_2}^{r_2}, \ldots, W_{m_n}^{r_n}\right]:=\frac{1}{2^{\alpha}}\bigg(\frac{[-2\sum_{l=1}^{n}m_{l}]_{\mathcal{R}(p,q)}}{[-\sum_{l=1}^{n}m_{l}]_{\mathcal{R}(p,q)}}\bigg)^{\alpha}\epsilon_{1 2 \cdots n}^{i_1 i_2 \cdots i_n}
W_{m_{i_l}}^{r_{i_l}}W_{m_{i_2}}^{r_{i_2}}\cdots W_{m_{i_n}}^{r_{i_n}},
\end{equation}		
where $\alpha=\frac{1+(-1)^n }{ 2},$  
and $\epsilon _{1\cdots p}^{i_{1}\cdots i_{p}}$  is the L\'evi-Civit\'a symbol defined by:
	\begin{eqnarray*}\label{LCs}
	\epsilon^{j_1 \cdots j_p}_{i_1 \cdots i_p}:= det\left( \begin{array} {ccc}
	\delta^{j_1}_{i_1} &\cdots&  \delta^{j_1}_{i_p}   \\ 
	\vdots && \vdots \\
	\delta^{j_p}_{i_1} & \cdots& \delta^{j_p}_{i_p}
	\end {array} \right) .
	\end{eqnarray*}

 Then, the $\mathcal{R}(p,q)$-$W_{1+\infty}$ $n$-algebra is governed by the operators \eqref{Rpqop1} with the commutation relation: 
 \begin{align}\label{RpqWnalg}
 [W_{m_1}^{r_1}, W_{m_2}^{r_2}, \ldots, W_{m_n}^{r_n}]
 &=\frac{K(P,Q)}{2^{\alpha}}\bigg(\frac{[-2\sum_{l=1}^{n}m_{l}]_{\mathcal{R}(p,q)}}{[-\sum_{l=1}^{n}m_{l}]_{\mathcal{R}(p,q)}}\bigg)^{\alpha}\epsilon_{1 2 \cdots n}^{i_1 i_2 \cdots i_n}\nonumber\\
 &\times \dsum_{\alpha_1=0}^{\beta_1}
 \dsum_{\alpha_2=0}^{\beta_2}
 \cdots \dsum_{\alpha_{n-1}=0}^{\beta_{n-1}}
 \mathbb{C}_{\beta_1}^{\alpha_1}\mathbb{C}_{\beta_2}^{\alpha_2}\cdots \mathbb{C}_{\beta_{n-1}}^{\alpha_{n-1}}
 \cdot \mathbb{A}_{m_{i_2}+r_{i_2}-1}^{\alpha_1}\nonumber\\
 &\times\mathbb{A}_{m_{i_3}+r_{i_3}-1}^{\alpha_2}\cdots \mathbb{A}_{m_{i_n}+r_{i_n}-1}^{\alpha_{n-1}}q^{\lambda}p^{\eta}
 W_{m_1+m_2+\cdots +m_{n}}^{r_1+\cdots+r_n-(n-1)-\alpha_1-\cdots-\alpha_{n-1}},
 \end{align}
 where
 $$\beta_k=\left\{
 \begin{array}{cc}
 r_{i_1}-1,& k=1, \\
 \dsum_{j=1}^k r_{i_j}-k-\dsum_{i=1}^{k-1}\alpha_i,   & 2\leqslant k\leqslant n-1,\\
 \end{array}\right. \quad  
 \eta=\dsum_{i=1}^{n}\alpha_i,$$
 and 
 	\begin{eqnarray*}
 	\lambda=\sum_{j=1}^{n
 	}\bigg(\beta_{j}-\sum_{j=1}^{n}\alpha_{j}\bigg)\big(m_{i_{j}+1}+r_{i_{j}+1}-1\big).
 	\end{eqnarray*}
 According to the associativity of the  $\mathcal{R}(p,q)$-operators $W_m^r$ \eqref{Rpqop1}, the $\mathcal{R}(p,q)$-$n$-algebra \eqref{Rpqn-bracket} with $n$ even
 satisfies the deformed Jacobi identity (GJI)
 \begin{equation}\label{ShJ}
 \epsilon _{12\cdots 2n-1}^{i_{1}i_{2}\cdots i_{2n-1}}\left[ [B_{i_{1}},B_{i_{2}}\cdots
 ,B_{i_{n}}],B_{i_{n+1}},\cdots ,B_{i_{2n-1}}\right] =0.
 \end{equation}
 Thus, it is easy to show  that the $\mathcal{R}(p,q)$-$W_{1+\infty}$ $n$-algebra \eqref{RpqWnalg} with $n$ even verify the deformed Jacobi identity \eqref{ShJ}. Then, we can conclude that  is a
 $\mathcal{R}(p,q)$-Lie algebra  or higher order $\mathcal{R}(p,q)$-Lie algebra.
 
Moreover, it is well-known that the $\mathcal{R}(p,q)$-Witt algebra is the unique nontrivial subalgebra of the $\mathcal{R}(p,q)$-$W_{1+\infty}$ algebra \eqref{Rpqwalg}.
Therefore it is convenient to investigate the subalgebra in \eqref{RpqWnalg}.
 
 We introduce the $2n$-commutator of the $\mathcal{R}(p,q)$-operators $ W_m^{n+1} $ for $n+1\geqslant 2$. Then, by scaling the $\mathcal{R}(p,q)$-operators $\tilde{W}_{m}^{n+1}\rightarrow Q^{-\frac{1}{2n-1}}W_{m}^{n+1}$,
 with
 \begin{eqnarray*}
 	Q=\mathbb{C}_{\beta_1}^{\alpha_1}
 	\mathbb{C}_{\beta_2}^{\alpha_2}\cdots  \mathbb{C}_{\beta_{2n-1}}^{\alpha_{2n-1}}
 	\epsilon_{1 2 \cdots {(2n-1)}}^{\alpha_1 \alpha_2 \cdots \alpha_{2n-1}},\end{eqnarray*}
 	 we obtain
 the $\mathcal{R}(p,q)$-sub-$2n$-algebra as follows:
 	\begin{align}\label{Rpqsub2nalg}
 	\left[\tilde{W}_{m_1}^{n+1}, \tilde{W}_{m_2}^{n+1}, \ldots , \tilde{W}_{m_{2n}}^{n+1}\right]&=\frac{K(P,Q
 		)}{2}\frac{[-2\sum_{l=1}^{2n}m_{l}]_{\mathcal{R}(p,q)}}{[-\sum_{l=1}^{2n}m_{l}]_{\mathcal{R}(p,q)}}\nonumber\\&\times\dprod_{1\leqslant j<k\leqslant 2n}\big([m_k+1]_{\mathcal{R}(p,q)}-[m_j+1]_{\mathcal{R}(p,q)}\big)
 	\tilde{W}_{m_1+m_2+\cdots +m_{2n}}^{n+1},
 	\end{align}
 	and for $k\geq 2n+1,$ we obtain:
 \begin{eqnarray*}
 \left[\tilde{W}_{m_1}^{n+1}, \ldots ,\tilde{W}_{m_{2n+1}}^{n+1}\right]=0.
 \end{eqnarray*}
 For the particular case to  $n=1$  in the relation \eqref{Rpqsub2nalg},  we deduce:
 \begin{eqnarray*}
 \left[\tilde{W}_{m_1}^{2}, \tilde{W}_{m_2}^{2}\right]=\frac{K(P,Q
 	)}{2}\frac{[-2\sum_{l=1}^{2}m_{l}]_{\mathcal{R}(p,q)}}{[-\sum_{l=1}^{2}m_{l}]_{\mathcal{R}(p,q)}}\big([m_2+1]_{\mathcal{R}(p,q)}-[m_1+1]_{\mathcal{R}(p,q)}\big)
 \tilde{W}_{m_1+m_2}^{2},
 \end{eqnarray*}
 \begin{example}
 	The $\mathcal{R}(p,q)$ Sub-4-algebra is derived as:
 	\begin{align*}
 	\left[\tilde{W}_{m_1}^3, \tilde{W}_{m_2}^3, \tilde{W}_{m_3}^3, \tilde{W}_{m_4}^3\right]
 	&=\frac{K(P,Q)}{16}\frac{[-2\sum_{l=1}^{4}m_{l}]_{\mathcal{R}(p,q)}}{[-\sum_{l=1}^{4}m_{l}]_{\mathcal{R}(p,q)}}\big([m_4]_{\mathcal{R}(p,q)}-[m_3]_{\mathcal{R}(p,q)}\big)\nonumber\\&\times\big([m_4]_{\mathcal{R}(p,q)}-[m_2]_{\mathcal{R}(p,q)}\big)\big([m_4]_{\mathcal{R}(p,q)}-[m_1]_{\mathcal{R}(p,q)} \big)\nonumber\\&\times\big([m_3]_{\mathcal{R}(p,q)}-[m_2]_{\mathcal{R}(p,q)}\big)
 	\big([m_3]_{\mathcal{R}(p,q)}-[m_1]_{\mathcal{R}(p,q)}\big)\nonumber\\&\times\big([m_2]_{\mathcal{R}(p,q)}-[m_1]_{\mathcal{R}(p,q)}\big) \tilde{W}_{m_1+m_2+m_3+m_4}^3.
 	\end{align*}
 	When $n\geqslant 5$, we have the null $n$-algebra
 	\begin{eqnarray*}
 	\left[\tilde{W}_{m_1}^3, \tilde{W}_{m_2}^3, \cdots, \tilde{W}_{m_n}^3\right]=0.
 	\end{eqnarray*}
 \end{example}
 
 Now let us discuss the central extensions of \eqref{Rpqsub2nalg}
 \begin{equation}\label{Rpqcsub2nalg}
 \left[\tilde{W}_{m_1}^{n+1}, \tilde{W}_{m_2}^{n+1}, \ldots , \tilde{W}_{m_{2n}}^{n+1}\right]=f(m_1,\cdots,m_{2n})\tilde{W}_{m_1+m_2+\cdots +m_{2n}}^{n+1}+C(m_1,m_2,\cdots,m_{2n}),
 \end{equation}
 where $f(m_1,\cdots,m_{2n})=\frac{K(P,Q)}{2}\frac{[-2\sum_{l=1}^{2n}m_{l}]_{\mathcal{R}(p,q)}}{[-\sum_{l=1}^{2n}m_{l}]_{\mathcal{R}(p,q)}}\dprod_{1\leqslant j<k\leqslant 2n}\big([m_k+1]_{\mathcal{R}(p,q)}-[m_j+1]_{\mathcal{R}(p,q)}\big)$ and $C(m_1,m_2,\cdots,m_{2n})$ is the central term to be determined.
 
 From the skewsymmetry of (\ref{Rpqcsub2nalg})
 and the GJI \eqref{ShJ}, we obtain the following relations of the
 central terms:
 \begin{eqnarray*}\label{CShJ1}
 &&C(m_1,\cdots,m_i,\cdots,m_j,\cdots,m_{2n})=-C(m_1,\cdots,m_j,\cdots,m_i,\cdots,m_{2n}),\end{eqnarray*}
 and
 \begin{equation}\label{CShJ2}
 \epsilon _{12\cdots 4n-1}^{i_{1}i_{2}\cdots i_{4n-1}}f(m_{i_{1}},m_{i_{2}},\cdots
 ,m_{i_{2n}})C(m_{i_1}+\cdots+m_{i_{2n}},m_{i_{2n+1}},\cdots ,m_{i_{4n-1}}) =0.
 \end{equation} 
 By cheking the relation \eqref{CShJ2},
 we  derive the following solution:
 \begin{align*}
 C(m_1,m_2,\cdots,m_{2n})&=\frac{c}{12 \times2^n n!}
 \epsilon_{1 2 \cdots {2n}}^{i_1 i_2 \cdots i_{2n}}\dprod_{k=1}^n\frac{p^{m_{i_{2k-1}}}\,[m_{i_{2k-1}}]_{\mathcal{R}(p,q)}}{q^{m_{i_{2k-1}}}[2m_{i_{2k-1}}]_{\mathcal{R}(p,q)}}\nonumber\\
 &\times[m_{i_{2k-1}}-1]_{\mathcal{R}(p,q)}[m_{i_{2k-1}}]_{\mathcal{R}(p,q)}[m_{i_{2k-1}}+1]_{\mathcal{R}(p,q)}\delta_{m_{i_{2k-1}}+m_{i_{2k}},0},
 \end{align*}
 where $c$ is an arbitrary constant.
 
 Thus, 
 	the $\mathcal{R}(p,q)$-$W_{1+\infty}$ sub-$2n$-algebra with central extension is determined as follows :
 	\begin{small}
 		\begin{align}\label{RpqcsubW2nalg} \left[\tilde{W}_{m_1}^{n+1}, \ldots , \tilde{W}_{m_{2n}}^{n+1}\right]_{\mathcal{R}(p,q)}&=\frac{K(P,Q)}{2}\frac{\left[-2\sum_{l=1}^{2n}m_{l}\right]_{\mathcal{R}(p,q)}}{\left[-\sum_{l=1}^{2n}m_{l}\right]_{\mathcal{R}(p,q)}}\dprod_{1\leqslant j<k\leqslant 2n}\big([m_k+1]_{\mathcal{R}(p,q)}-[m_j+1]_{\mathcal{R}(p,q)}\big)
 	\tilde{W}_{\sum_{i=1}^{2n}m_i}^{n+1}\nonumber\\&+\frac{c}{6\times 2^n\,n!}
 	\epsilon_{1 2 \cdots {2n}}^{i_1 i_2 \cdots i_{2n}}
 	\dprod_{k=1}^n\bigg(\frac{p^{m_{i_{2k-1}}}[m_{i_{2k-1}}]_{\mathcal{R}(p,q)}}{q^{m_{i_{2k-1}}}[2m_{i_{2k-1}}]_{\mathcal{R}(p,q)}}[m_{i_{2k-1}}-1]_{\mathcal{R}(p,q)}\nonumber\\
 	&\times[m_{i_{2k-1}}]_{\mathcal{R}(p,q)}[m_{i_{2k-1}}+1]_{\mathcal{R}(p,q)}\delta_{m_{i_{2k-1}}+m_{i_{2k}},0}\bigg),
 	\end{align}
 \end{small}
 	where $c$ is an arbitrary constant.
 \begin{example}
 	As  example, taking the $\mathcal{R}(p,q)$-operator $\tilde{W}_{m}^2$ in \eqref{RpqcsubW2nalg},
 	we obtain the $\mathcal{R}(p,q)$-Virasoro algebra
 	\begin{align*}
 	\left[\tilde{W}_{m_1}^2,\tilde{W}_{m_2}^2\right]_{\mathcal{R}(p,q)}
 	&=\frac{K(P,Q)}{2}\frac{\left[-2\sum_{l=1}^{2}m_{l}\right]_{\mathcal{R}(p,q)}}{\left[-\sum_{l=1}^{2}m_{l}\right]_{\mathcal{R}(p,q)}}\big([m_2+1]_{\mathcal{R}(p,q)}-[m_1+1]_{\mathcal{R}(p,q)}\big) \tilde{W}_{m_1+m_2}^2\nonumber\\
 	&+\frac{c}{12}\frac{p^{m_{1}}[m_{1}]_{\mathcal{R}(p,q)}}{q^{m_{1}}[2m_{1}]_{\mathcal{R}(p,q)}}[m_{1}-1]_{\mathcal{R}(p,q)}[m_{1}]_{\mathcal{R}(p,q)}[m_{1}+1]_{\mathcal{R}(p,q)}\delta_{m_1+m_2,0}.
 	\end{align*}
 \end{example}
\begin{remark}
	Particular cases of $W_{1+\infty}$ $n$-algebra from quantum algebras are deduced as follows:
	\begin{enumerate}
		\item[(i)] 
	Taking $\mathcal{R}(x)=\frac{x-1}{q-1},$ we deduce the $W_{1+\infty}$ $n$-algebra from the $q$-deformed quantum algebra \cite{AC}. Then, the $q$-deformed Leibniz formula is given by: 
\begin{eqnarray*}
\mathcal{D}_{q}^r\big(f(x)g(x)\big)&=&\sum_{j=0}^{r}\bigg[\begin{array}{c} r  \\ j\end{array} \bigg]_{q}\mathcal{D}_{q}^{j}\big(f(xq^{r-j})\big)\mathcal{D}_{q}^{r-j}g(x)\label{qRL},
\end{eqnarray*}
where for $0<q<1$ and $n\in\mathbb{N}$, the $q$-derivative, $q$-numbers  and $q$ binomial coefficients are:
\begin{equation*}
\mathcal{D}_{q}f(x)=\frac{f(x)-f(qx)}{(1-q)x},\quad [n]_q=\frac{1-q^n}{1-q},
\end{equation*}
and 
\begin{equation*}
    \bigg[\begin{array}{c} r  \\ j\end{array} \bigg]_{q}=\frac{[r]_q!}{[j]_q![r-j]_q!}.
\end{equation*}
The $q$-deformed operators
\begin{equation}\label{eq:walgoperator}
W_m^r=z^{m+r-1}\mathcal{D}_{q}^{r-1}, \ \ r\in \mathbb{Z}_+,\, m \in\mathbb{N}.
\end{equation}
satisfies the commutation relation:
\begin{align}\label{eq:walgebra}
\left[W_{m}^{r}, W_{n}^{s}\right]
&=\bigg(\dsum_{j=0}^{r-1}q^{(r-1-j)(n+s-1)}\mathbb{C}_{r-1}^{j}\mathbb{A}_{n+s-1}^{j}\nonumber\\&-\dsum_{j=0}^{s-1}q^{(s-1-j)(m+r-1)}\mathbb{C}_{s-1}^{j}
\mathbb{A}_{m+r-1}^{j}\bigg)W_{m+n}^{r+s-1-j},
\end{align}
\begin{eqnarray*}
\mathbb{C}^k_n:= \frac{[n]_q!}{[k]_q![n-k]_q!}, n\geq k,\,\mbox{and}\, \mathbb{A}_n^{k}=\left\{\begin{array}{cc}
[n]_{q}[n-1]_{q}\cdots[n-k+1]_{q},& k\leqslant n,\\
0,                  &k>n.\end{array}\right.
\end{eqnarray*}
	Note that, by taking $r=s=2$ in the relation \eqref{eq:walgebra}, we obtain the $q$-algebra:
	\begin{equation}\label{qWalg}
	\left[W_{m}^2, W_{n}^2\right]=\big(q^{n+1}-q^{m+1}\big)W_{m+n}^3+\big([n+1]_{q}-[m+1]_{\mathcal{R}(q)}\big)W_{m+n}^2
	\end{equation}
and when $q\longrightarrow 1$ in the relation \eqref{qWalg}, we recovered the  Witt algebra:\begin{equation*}
		\left[W_{m}^2, W_{n}^2\right]=\big(n-m\big)W_{m+n}^2.
		\end{equation*}
Moreover, the $q$-$W_{1+\infty}$ $n$-algebra is generated by the $n$-commutators: 
\begin{align}\label{qWnalg}
[W_{m_1}^{r_1}, W_{m_2}^{r_2}, \ldots, W_{m_n}^{r_n}]
&=\frac{1}{2^{\alpha}}\bigg(\frac{[-2\sum_{l=1}^{n}m_{l}]_{q}}{[-\sum_{l=1}^{n}m_{l}]_{q}}\bigg)^{\alpha}\epsilon_{1 2 \cdots n}^{i_1 i_2 \cdots i_n}\nonumber\\
&\times \dsum_{\alpha_1=0}^{\beta_1}
\dsum_{\alpha_2=0}^{\beta_2}
\cdots \dsum_{\alpha_{n-1}=0}^{\beta_{n-1}}
\mathbb{C}_{\beta_1}^{\alpha_1}\mathbb{C}_{\beta_2}^{\alpha_2}\cdots \mathbb{C}_{\beta_{n-1}}^{\alpha_{n-1}}
\cdot \mathbb{A}_{m_{i_2}+r_{i_2}-1}^{\alpha_1}\nonumber\\
&\times\mathbb{A}_{m_{i_3}+r_{i_3}-1}^{\alpha_2}\cdots \mathbb{A}_{m_{i_n}+r_{i_n}-1}^{\alpha_{n-1}}
q^{\lambda}W_{m_1+m_2+\cdots +m_{n}}^{r_1+\cdots+r_n-(n-1)-\alpha_1-\cdots-\alpha_{n-1}},
\end{align}
where
\begin{eqnarray*}\beta_k=\left\{
\begin{array}{cc}
r_{i_1}-1,& k=1, \\
\dsum_{j=1}^k r_{i_j}-k-\dsum_{i=1}^{k-1}\alpha_i,   & 2\leqslant k\leqslant n-1\\
\end{array}\right.
\end{eqnarray*}
and
\begin{equation*} 	\lambda=\sum_{j=1}^{n
}\bigg(\beta_{j}-\sum_{j=1}^{n}\alpha_{j}\bigg)\big(m_{i_{j}+1}+r_{i_{j}+1}-1\big).
\end{equation*}
According to the associativity of the  $q$-operators $W_m^r$ \eqref{eq:walgoperator},
the $q$-$n$-algebra \eqref{qWnalg} with $n$ even
satisfies the deformed Jacobi identity (GJI)\eqref{ShJ}
Hence, we can concluded that the $q$-$W_{1+\infty}$ $n$-algebra \eqref{qWnalg} with $n$ even is a
$q$-Lie algebra  or higher order $q$-Lie algebra.

Furthermore, the $q$-sub-$2n$-algebra is derived as:
\begin{align*}
\left[\tilde{W}_{m_1}^{n+1}, \tilde{W}_{m_2}^{n+1}, \ldots , \tilde{W}_{m_{2n}}^{n+1}\right]&=\frac{[-2\sum_{l=1}^{2n}m_{l}]_{q}}{2[-\sum_{l=1}^{2n}m_{l}]_{q}}\dprod_{1\leqslant j<k\leqslant 2n}\big([m_k]_q-[m_j]_{q}\big)
\tilde{W}_{m_1+m_2+\cdots +m_{2n}}^{n+1}\nonumber\\
\,\left[\tilde{W}_{m_1}^{n+1}, \ldots ,\tilde{W}_{m_{2n+1}}^{n+1}\right]&=0,\quad k\geq 2n+1.
\end{align*} 
Taking  $n=1$ in the above relation,
  we obtain
the $q$-algebra
 \begin{eqnarray*}
	\left[\tilde{W}_{m_1}^{2}, \tilde{W}_{m_2}^{2}\right]=\frac{1+q^{-m_1}}{2}\big([m_2]_q-[m_1]_{q}\big)
	\tilde{W}_{m_1+m_2}^{2}.
	\end{eqnarray*}
For example, we have the  $q$-sub-$4$-algebras:
	\begin{align*}
	\left[\tilde{W}_{m_1}^3, \tilde{W}_{m_2}^3, \tilde{W}_{m_3}^3, \tilde{W}_{m_4}^3\right]_q
	&=\frac{[-2\sum_{l=1}^{4}m_{l}]_{q}}{16[-\sum_{l=1}^{4}m_{l}]_{q}}\big([m_4]_q-[m_3]_{q}\big)\big([m_4]_q-[m_2]_{q}\big)\nonumber\\&\times\big([m_4]_q-[m_1]_{q}\big) \big([m_3]_q-[m_2]_{q}\big)
	\nonumber\\&\times\big([m_3]_q-[m_1]_{q}\big)\big([m_2]_q-[m_1]_{q}\big)\tilde{W}_{m_1+m_2+m_3+m_4}^3\nonumber\\
	\,\left[\tilde{W}_{m_1}^3, \tilde{W}_{m_2}^3, \cdots, \tilde{W}_{m_n}^3\right]&=0,\quad n\geqslant 5. 
	\end{align*}
Thus, we obtain the $q$-$W_{1+\infty}$ sub-$2n$-algebra with central extension given by:
\begin{align*} \left[\tilde{W}_{m_1}^{n+1}, \ldots , \tilde{W}_{m_{2n}}^{n+1}\right]&=\frac{[-2\sum_{l=1}^{2n}m_{l}]_{q}}{2[-\sum_{l=1}^{2n}m_{l}]_{q}}\dprod_{1\leqslant j<k\leqslant 2n}\big([m_k]_q-[m_j]_{q}\big)
\tilde{W}_{\sum_{i=1}^{2n}m_i}^{n+1}+\frac{c}{6\times 2^n\,n!}
\nonumber\\&\times\epsilon_{1 2 \cdots {2n}}^{i_1 i_2 \cdots i_{2n}}
\dprod_{k=1}^n\frac{[m_{i_{2k-1}}]_{q}}{q^{m_{i_{2k-1}}}}[2m_{i_{2k-1}}]_{q}
[m_{i_{2k-1}}-1]_{q}\nonumber\\&\times[m_{i_{2k-1}}]_{q}[m_{i_{2k-1}}+1]_{q}\delta_{m_{i_{2k-1}}+m_{i_{2k}},0},
\end{align*}
where $c$ is an arbitrary constant.
	As the example,
	we have the $q$-Virasoro algebra
	\begin{align*}
	\left[\tilde{W}_{m_1}^2,\tilde{W}_{m_2}^2\right]
	&=\big([m_2]_q-[m_1]_{q}\big)\tilde{W}_{m_1+m_2}^2\\&+\frac{c}{12}\frac{[m_{1}]_{q}}{q^{m_{1}}[2m_{1}]_{q}}[m_{1}-1]_{q}[m_{1}]_{q}[m_{1}+1]_{q}\delta_{m_1+m_2,0}.
	\end{align*}
\item[(ii)] Setting $\mathcal{R}(x,y)=\frac{x-y}{p-q}$, we obtain:
the $(p,q)$-Leibniz formula is given by:
\begin{equation*}
\mathcal{D}^r_{p,q}\big(f(x)g(x)\big)=\sum_{j=0}^{r}\bigg[\begin{array}{c} r  \\ j\end{array} \bigg]_{p,q}\mathcal{D}^{j}_{p,q}\big(f(xq^{r-j})\big)\mathcal{D}^{r-j}_{p,q}g\big(xp^{j}\big),
\end{equation*}
where for $0<q<p<1$ and $n\in\mathbb{N}$, the $(p,q)$-derivative, $(p,q)$-numbers  and $(p,q)$ binomial coefficients are:
\begin{equation*}
\mathcal{D}_{p,q}f(x)=\frac{f(px)-f(qx)}{(p-q)x},\quad [n]_{p,q}=\frac{p^n-q^n}{p-q},
\end{equation*}
and 
\begin{equation*}
    \bigg[\begin{array}{c} r  \\ j\end{array} \bigg]_{p,q}=\frac{[r]_{p,q}!}{[j]_{p,q}![r-j]_{p,q}!}.
\end{equation*}
Moreover,  the $(p,q)$-operators
\begin{equation*}
W_m^r=z^{m+r-1}\mathcal{D}_{p,q}^{r-1}, \ \ r\in \mathbb{Z}_+,\, m \in\mathbb{N},
\end{equation*}
  verify the commutation relation:
	\begin{align*}
\left[W_{m}^{r}, W_{n}^{s}\right]
&=\bigg(\dsum_{j=0}^{r-1}q^{(r-1-j)(n+s-1)}p^{j}\mathbb{C}_{r-1}^{j}\mathbb{A}_{n+s-1}^{j}\\&-\dsum_{j=0}^{s-1}q^{(s-1-j)(m+r-1)}p^{j}\mathbb{C}_{s-1}^{j}
\mathbb{A}_{m+r-1}^{j}\bigg)W_{m+n}^{r+s-1-j},
\end{align*}
where  
$$\mathbb{A}_n^{k}=\left\{\begin{array}{cc}
[n]_{p,q}[n-1]_{p,q}\cdots[n-k+1]_{p,q},& k\leqslant n,\\
0,                  &k>n.\end{array}\right. $$
and 
\begin{equation*}
\mathbb{C}^k_n:= \frac{[n]_{p,q}!}{[k]_{p,q}![n-k]_{p,q}!}, \quad n\geq k.
\end{equation*}
By setting $r=s=2$, we obtain the $(p,q)$-algebra:
	\begin{eqnarray*}
	\left[W_{m}^2, W_{n}^2\right]=\big(q^{n+1}-q^{m+1}\big)W_{m+n}^3+\big([n+1]_{p,q}-[m+1]_{p,q}\big)W_{m+n}^2.
	\end{eqnarray*} 
Besides, the $(p,q)$-$W_{1+\infty}$ $n$-algebra is given the commutation relation: 
\begin{align*}
[W_{m_1}^{r_1}, W_{m_2}^{r_2}, \ldots, W_{m_n}^{r_n}]
&=\frac{1}{2^{\alpha}}\bigg(\frac{[-2\sum_{l=1}^{n}m_{l}]_{p,q}}{[-\sum_{l=1}^{n}m_{l}]_{p,q}}\bigg)^{\alpha}\epsilon_{1 2 \cdots n}^{i_1 i_2 \cdots i_n}\nonumber\\
&\times \dsum_{\alpha_1=0}^{\beta_1}
\dsum_{\alpha_2=0}^{\beta_2}
\cdots \dsum_{\alpha_{n-1}=0}^{\beta_{n-1}}
\mathbb{C}_{\beta_1}^{\alpha_1}\mathbb{C}_{\beta_2}^{\alpha_2}\cdots \mathbb{C}_{\beta_{n-1}}^{\alpha_{n-1}}
\cdot \mathbb{A}_{m_{i_2}+r_{i_2}-1}^{\alpha_1}\nonumber\\
&\times\mathbb{A}_{m_{i_3}+r_{i_3}-1}^{\alpha_2}\cdots \mathbb{A}_{m_{i_n}+r_{i_n}-1}^{\alpha_{n-1}}q^{\lambda}p^{\eta}
W_{m_1+m_2+\cdots +m_{n}}^{r_1+\cdots+r_n-(n-1)-\alpha_1-\cdots-\alpha_{n-1}},
\end{align*}
where
$$\beta_k=\left\{
\begin{array}{cc}
r_{i_1}-1,& k=1, \\
\dsum_{j=1}^k r_{i_j}-k-\dsum_{i=1}^{k-1}\alpha_i,   & 2\leqslant k\leqslant n-1,\\
\end{array}\right. \quad  
\eta=\dsum_{i=1}^{n}\alpha_i,$$
and 
\begin{eqnarray*}
\lambda=\sum_{j=1}^{n
}\bigg(\beta_{j}-\sum_{j=1}^{n}\alpha_{j}\bigg)\big(m_{i_{j}+1}+r_{i_{j}+1}-1\big).
\end{eqnarray*}
The $(p,q)$-$n$-algebra with $n$ even
satisfies the deformed Jacobi identity (GJI).
Thus, the $(p,q)$-$W_{1+\infty}$ $n$-algebra with $n$ even is a
$(p,q)$-Lie algebra  or higher order $(p,q)$-Lie algebra.

Moreover, the $(p,q)$-sub-$2n$-algebra is given:
\begin{align*}
\,\left[\tilde{W}_{m_1}^{n+1}, \tilde{W}_{m_2}^{n+1}, \ldots , \tilde{W}_{m_{2n}}^{n+1}\right]&=\frac{1}{2}\frac{[-2\sum_{l=1}^{2n}m_{l}]_{p,q}}{[-\sum_{l=1}^{2n}m_{l}]_{p,q}}\nonumber\\&\times\dprod_{1\leqslant j<k\leqslant 2n}\big([m_k+1]_{p,q}-[m_j+1]_{p,q}\big)
\tilde{W}_{m_1+m_2+\cdots +m_{2n}}^{n+1},\nonumber\\
\,\left[\tilde{W}_{m_1}^{n+1}, \ldots ,\tilde{W}_{m_{2n+1}}^{n+1}\right]&=0,\quad k\geq 2n+1.
\end{align*}
For  $n=1$,  we deduce:
\begin{eqnarray*}
\left[\tilde{W}_{m_1}^{2}, \tilde{W}_{m_2}^{2}\right]=\frac{1}{2}\frac{[-2\sum_{l=1}^{2}m_{l}]_{p,q}}{[-\sum_{l=1}^{2}m_{l}]_{p,q}}\big([m_2+1]_{p,q}-[m_1+1]_{p,q}\big)
\tilde{W}_{m_1+m_2}^{2},
\end{eqnarray*}
	The $(p,q)$-Sub-4-algebra is derived as:
	\begin{align*}
	\,\left[\tilde{W}_{m_1}^3, \tilde{W}_{m_2}^3, \tilde{W}_{m_3}^3, \tilde{W}_{m_4}^3\right]
	&=\frac{1}{16}\frac{[-2\sum_{l=1}^{4}m_{l}]_{p,q}}{[-\sum_{l=1}^{4}m_{l}]_{p,q}}\big([m_4]_{p,q}-[m_3]_{p,q}\big)\big([m_4]_{p,q}-[m_2]_{p,q}\big)\nonumber\\&\times\big([m_4]_{p,q}-[m_1]_{p,q}\big) \big([m_3]_{p,q}-[m_2]_{p,q}\big)\nonumber\\&\times
	\big([m_3]_{p,q}-[m_1]_{p,q}\big)\big([m_2]_{p,q}-[m_1]_{p,q}\big) \tilde{W}_{m_1+m_2+m_3+m_4}^3,\nonumber\\ 
	\,\left[\tilde{W}_{m_1}^3, \tilde{W}_{m_2}^3, \cdots, \tilde{W}_{m_n}^3\right]&=0,\quad n\geqslant 5.
	\end{align*}
Thus, 
the $(p,q)$-$W_{1+\infty}$ sub-$2n$-algebra with central extension is determined as follows :
	\begin{align*} \left[\tilde{W}_{m_1}^{n+1}, \ldots , \tilde{W}_{m_{2n}}^{n+1}\right]_{p,q}&=\frac{1}{2}\frac{\left[-2\sum_{l=1}^{2n}m_{l}\right]_{p,q}}{\left[-\sum_{l=1}^{2n}m_{l}\right]_{p,q}}\dprod_{1\leqslant j<k\leqslant 2n}\big([m_k+1]_{p,q}-[m_j+1]_{p,q}\big)
	\tilde{W}_{\sum_{i=1}^{2n}m_i}^{n+1}\nonumber\\&+\frac{c}{6\times 2^n\,n!}
	\epsilon_{1 2 \cdots {2n}}^{i_1 i_2 \cdots i_{2n}}
	\dprod_{k=1}^n\bigg(\frac{p^{m_{i_{2k-1}}}[m_{i_{2k-1}}]_{p,q}}{q^{m_{i_{2k-1}}}[2m_{i_{2k-1}}]_{p,q}}[m_{i_{2k-1}}-1]_{p,q}\nonumber\\
	&\times[m_{i_{2k-1}}]_{p,q}[m_{i_{2k-1}}+1]_{p,q}\delta_{m_{i_{2k-1}}+m_{i_{2k}},0}\bigg),
	\end{align*}
where $c$ is an arbitrary constant.
	As  example,
	we get the $(p,q)$-Virasoro algebra
	\begin{align*}
	\left[\tilde{W}_{m_1}^2,\tilde{W}_{m_2}^2\right]_{p,q}
	&=\frac{1}{2}\frac{\left[-2\sum_{l=1}^{2}m_{l}\right]_{p,q}}{\left[-\sum_{l=1}^{2}m_{l}\right]_{p,q}}\big([m_2+1]_{p,q}-[m_1+1]_{p,q}\big)\tilde{W}_{m_1+m_2}^2\nonumber\\
	&+\frac{c}{12}\frac{p^{m_{1}}[m_{1}]_{p,q}}{q^{m_{1}}[2m_{1}]_{p,q}}[m_{1}-1]_{p,q}[m_{1}]_{p,q}[m_{1}+1]_{p,q}\delta_{m_1+m_2,0}.
	\end{align*}
\end{enumerate}
\end{remark}
\section{Another $\mathcal{R}(p,q)$-$W_{1+\infty}$-$n$-algebra} In this section, we  construct the $W_{1+\infty}$ $n$-algebra from the quantum algebra \cite{HB1}. Particular cases are also deduced. Then, let us introduce the $\mathcal{R}(p,q)$-operators $\mathcal{W}^s_m$
in the following
way,
\begin{equation}\label{Rpqa1}
\mathcal{W}^s_m=\frac{1}{2(m+s)}[x^2,\, \mathcal{W}^{s-1}_{m+2} ],\quad s=2,3, \cdots,
\end{equation}
and we set
\begin{equation}\label{Rpqa2}
\mathcal{W}^1_m=-D^m_{\mathcal{R}(p,q)}, \quad m\geq 1.
\end{equation}
Substituting the $\mathcal{R}(p,q)$-operators \eqref{Rpqa2} into the recursive relation \eqref{Rpqa1}, we obtain:
\begin{equation*}
\mathcal{W}^s_m=(-1)^s\sum_{k=0}^{s-1}\frac{\mathbb{C}^k_{s-1}\,\mathbb{A}^k_{m+s-1}}{2^k}\,x^{s-1-k}\,D_{\mathcal{R}(p,q)}^{m+s-1-k}.
\end{equation*}
Moreover, they  form the $\mathcal{R}(p,q)$-$W_{1+\infty}$ algebra: 
\begin{equation}\label{Rqa4}
\left[\mathcal{W}^s_n,\,\mathcal{W}^r_m\right]=\big([n(r-1)]_{\mathcal{R}(p,q)}-[m(s-1)]_{\mathcal{R}(p,q)}\big)\,\mathcal{W}^{s+r-2}_{n+m}+\cdots, 
\end{equation}
where $(\cdots)$ can be considered as the lower order
terms associated to the quantum effect. The first commutators of the relation \eqref{Rqa4} are given by:
\begin{align*}
\,\left[\mathcal{W}^1_n,\,\mathcal{W}^1_m\right]&=0,\quad \left[\mathcal{W}^2_n,\,\mathcal{W}^1_m\right]=[-m]_{\mathcal{R}(p,q)}\,\mathcal{W}^{1}_{n+m}\\
\,\left[\mathcal{W}^2_n,\,\mathcal{W}^2_m\right]&=\big([n]_{\mathcal{R}(p,q)}-[m]_{\mathcal{R}(p,q)}\big)\,\mathcal{W}^{2}_{n+m}
,\quad \left[\mathcal{W}^3_n,\,\mathcal{W}^1_m\right]=-[2m]_{\mathcal{R}(p,q)}\,\mathcal{W}^{2}_{n+m}\\
\,\left[\mathcal{W}^3_n,\,\mathcal{W}^2_m\right]&=\big([n]_{\mathcal{R}(p,q)}-[2m]_{\mathcal{R}(p,q)}\big)\,\mathcal{W}^{3}_{n+m}\nonumber\\&+\frac{1}{4}[n+2]_{\mathcal{R}(p,q)}[m+1]_{\mathcal{R}(p,q)}[m]_{\mathcal{R}(p,q)}\,\mathcal{W}^{1}_{n+m}\\
\,\left[\mathcal{W}^3_n,\,\mathcal{W}^3_m\right]&=\big([2n]_{\mathcal{R}(p,q)}-[2m]_{\mathcal{R}(p,q)}\big)\,\mathcal{W}^{4}_{n+m}\nonumber\\&- \frac{1}{2}\big([n]_{\mathcal{R}(p,q)}-[m]_{\mathcal{R}(p,q)}\big)[n+2]_{\mathcal{R}(p,q)}[m+2]_{\mathcal{R}(p,q)}\,\mathcal{W}^{2}_{n+m}.
\end{align*}
\begin{remark}
	It is necessary to furnish the $W_{1+\infty}$ algebra from some quantum algebras.
	\begin{enumerate}
		\item[(i)] The $q$-operators \begin{equation*}
		\mathcal{W}^s_m=(-1)^s\sum_{k=0}^{s-1}\frac{\mathbb{C}^k_{s-1}\,\mathbb{A}^k_{m+s-1}}{2^k}\,x^{s-1-k}\,D_{q}^{m+s-1-k}.
		\end{equation*}
		obey the $q$-$W_{1+\infty}$ algebra: 
		\begin{eqnarray*}
		\left[\mathcal{W}^s_n,\,\mathcal{W}^r_m\right]=\big([n(r-1)]_{\mathcal{R}(p,q)}-[m(s-1)]_{\mathcal{R}(p,q)}\big)\,\mathcal{W}^{s+r-2}_{n+m}+\cdots, 
		\end{eqnarray*}
		and the first commutators 
		\begin{align*}
		\,\left[\mathcal{W}^1_n,\,\mathcal{W}^1_m\right]&=0,\quad \left[\mathcal{W}^2_n,\,\mathcal{W}^1_m\right]=[-m]_{q}\,\mathcal{W}^{1}_{n+m}\\
		\,\left[\mathcal{W}^2_n,\,\mathcal{W}^2_m\right]&=\big([n]_{q}-[m]_{q}\big)\,\mathcal{W}^{2}_{n+m}
		,\quad \left[\mathcal{W}^3_n,\,\mathcal{W}^1_m\right]=-[2m]_{q}\,\mathcal{W}^{2}_{n+m}\\
		\,\left[\mathcal{W}^3_n,\,\mathcal{W}^2_m\right]&=\big([n]_{q}-[2m]_{q}\big)\,\mathcal{W}^{3}_{n+m}+\frac{1}{4}[n+2]_{q}[m+1]_{q}[m]_{q}\,\mathcal{W}^{1}_{n+m}\\
		\,\left[\mathcal{W}^3_n,\,\mathcal{W}^3_m\right]&=\big([2n]_{q}-[2m]_{q}\big)\,\mathcal{W}^{4}_{n+m}- \frac{1}{2}\big([n]_{q}-[m]_{q}\big)[n+2]_{q}[m+2]_{q}\,\mathcal{W}^{2}_{n+m}.
		\end{align*}
		\item[(ii)]The $(p,q)$-operators \begin{eqnarray*}
		\mathcal{W}^s_m=(-1)^s\sum_{k=0}^{s-1}\frac{\mathbb{C}^k_{s-1}\,\mathbb{A}^k_{m+s-1}}{2^k}\,x^{s-1-k}\,D_{p,q}^{m+s-1-k}
		\end{eqnarray*}
		generated the $(p,q)$-$W_{1+\infty}$ algebra: 
		\begin{eqnarray*}
		\left[\mathcal{W}^s_n,\,\mathcal{W}^r_m\right]=\big([n(r-1)]_{p,q}-[m(s-1)]_{p,q}\big)\,\mathcal{W}^{s+r-2}_{n+m}+\cdots, 
		\end{eqnarray*}
		with the first commutators : 
		\begin{align*}
		\,\left[\mathcal{W}^1_n,\,\mathcal{W}^1_m\right]&=0,\quad \left[\mathcal{W}^2_n,\,\mathcal{W}^1_m\right]=[-m]_{p,q}\,\mathcal{W}^{1}_{n+m}\\
		\,\left[\mathcal{W}^2_n,\,\mathcal{W}^2_m\right]&=\big([n]_{p,q}-[m]_{p,q}\big)\,\mathcal{W}^{2}_{n+m}
		,\quad \left[\mathcal{W}^3_n,\,\mathcal{W}^1_m\right]=[-2m]_{p,q}\,\mathcal{W}^{2}_{n+m}\\
		\,\left[\mathcal{W}^3_n,\,\mathcal{W}^2_m\right]&=\big([n]_{p,q}-[2m]_{p,q}\big)\,\mathcal{W}^{3}_{n+m}+\frac{1}{4}[n+2]_{p,q}[m+1]_{p,q}[m]_{p,q}\,\mathcal{W}^{1}_{n+m}\\
		\,\left[\mathcal{W}^3_n,\,\mathcal{W}^3_m\right]&=\big([2n]_{p,q}-[2m]_{p,q}\big)\,\mathcal{W}^{4}_{n+m}- \frac{1}{2}\big([n]_{p,q}-[m]_{p,q}\big)[n+2]_{p,q}[m+2]_{p,q}\,\mathcal{W}^{2}_{n+m}.
		\end{align*}
	\end{enumerate}
\end{remark}
As the case of the well-known operators $L_m=-x^{m+1}\,\mathcal{D}_{\mathcal{R}(p,q)}$, we note that the $\mathcal{R}(p,q)$-operators
\begin{equation*}
\mathcal{W}^2_m=\frac{1}{2(m+2)}[x^2,\, \mathcal{W}^{1}_{m+2} ]=x\,\mathcal{D}_{\mathcal{R}(p,q)}^{m+1}+ \frac{m+1}{2}\,\mathcal{D}_{\mathcal{R}(p,q)}^m,
\end{equation*}
also generated the $\mathcal{R}(p,q)$-Virasoro-Witt algebra. To investigate the difference of
these two realizations of deformed Virasoro-Witt algebra, we consider the case of the $\mathcal{R}(p,q)$-$3$-algebra.

The operator Nambu $3$-bracket is defined by \cite{Nambu,Takhtajan}:
\begin{align}\label{Rpqa7}
[A_1,A_2,A_3]&=A_1[A_2,A_3]+ A_2[A_3,A_1]+ A_3[A_1,A_2]\nonumber\\&=[A_2,A_3]A_1+ [A_3,A_1]A_2+ [A_1,A_2]A_3,
\end{align}
where the commutator is given by $[A_1,A_2]=A_1A_2-A_2A_1.$

By direct computation, the Nambu $\mathcal{R}(p,q)$-$3$-bracket $ [L_k,L_n,L_m]$ gives the null deformed $3$-algebra:
\begin{equation*}
[L_k,L_n,L_m]=0.
\end{equation*} 
Introducing the $\mathcal{R}(p,q)$-operators $\mathcal{W}^2_m$ and $\mathcal{W}^1_m$ into the Nambu $3$-brackets \eqref{Rpqa7}, we obtain the $\mathcal{R}(p,q)$-$3$-algbra:
\begin{align}
\, \left[\mathcal{W}^2_{m_1},\mathcal{W}^2_{m_2},\mathcal{W}^2_{m_3}\right]&=\frac{1}{4}\big([m_3]_{\mathcal{R}(p,q)}-[m_2]_{\mathcal{R}(p,q)}\big)\big([m_3]_{\mathcal{R}(p,q)}-[m_1]_{\mathcal{R}(p,q)}\big)\nonumber\\&\times\big([m_2]_{\mathcal{R}(p,q)}-[m_1]_{\mathcal{R}(p,q)}\big)\,\mathcal{W}^1_{m_1+m_2+m_3},\label{Rpqa9a}\\
\left[\mathcal{W}^2_{m_1},\mathcal{W}^2_{m_2},\mathcal{W}^1_{m_3}\right]&=\big([m_2]_{\mathcal{R}(p,q)}-[m_1]_{\mathcal{R}(p,q)}\big)\,\mathcal{W}^2_{m_1+m_2+m_3},\nonumber\\
\left[\mathcal{W}^2_{m_1},\mathcal{W}^1_{m_2},\mathcal{W}^1_{m_3}\right]&=\big([m_2]_{\mathcal{R}(p,q)}-[m_3]_{\mathcal{R}(p,q)}\big)\,\mathcal{W}^1_{m_1+m_2+m_3},\nonumber\\
\left[\mathcal{W}^2_{m_1},\mathcal{W}^1_{m_2},\mathcal{W}^1_{m_3}\right]&=0\label{Rpqa9d}.
\end{align}
It satisfies the Bremner identity (BI)\cite{Bremner1,Bremner2}:
\begin{align*}
\epsilon^{i_1i_2\cdots i_6}\left[\left[A\left[B_{i_1},B_{i_2},B_{i_3}\right],B_{i_4}\right],B_{i_5},B_{i_6}\right]&=\epsilon^{i_1i_2\cdots i_6}\left[\left[A,B_{i_1},B_{i_2}\right],\left[B_{i_3},B_{i_4},B_{i_5}\right],B_{i_6}\right],
\end{align*}
where $i_1,\cdots, i_6$ are implicitly summed from $1$ to $6$. Furthermore, the Filippov condition or fundamental identity
(FI),\cite{Filippov}
\begin{align}\label{Rpqa11}
\left[A,B,\left[C,D,E\right]\right]= \left[\left[A,B,C\right],D,E\right]+ \left[C,\left[A,B,D\right],E\right]+\left[C,D,\left[A,B,E\right]\right] 
\end{align}
doen't hold. Thus, we deduce that the $\mathcal{R}(p,q)$-$3$-algbra \eqref{Rpqa9a} to \eqref{Rpqa9d} is not
a deformed Nambu $3$-algebra.
\begin{remark}
	Particular cases of deformed $3$-algebra can be deduced as:
	\begin{enumerate}
		\item[(i)]the $q$-deformed $3$-algebra is given by the commutators:
		\begin{align*}
		 \left[\mathcal{W}^2_{m_1},\mathcal{W}^2_{m_2},\mathcal{W}^2_{m_3}\right]&=\frac{1}{4}\big([m_3]_{q}-[m_2]_{q}\big)\big([m_3]_{q}-[m_1]_{q}\big)\big([m_2]_{q}-[m_1]_{q}\big)\,\mathcal{W}^1_{m_1+m_2+m_3},\\
		\left[\mathcal{W}^2_{m_1},\mathcal{W}^2_{m_2},\mathcal{W}^1_{m_3}\right]&=\big([m_2]_{q}-[m_1]_{q}\big)\,\mathcal{W}^2_{m_1+m_2+m_3},	\\	\left[\mathcal{W}^2_{m_1},\mathcal{W}^1_{m_2},\mathcal{W}^1_{m_3}\right]&=\big([m_2]_{q}-[m_3]_{q}\big)\,\mathcal{W}^1_{m_1+m_2+m_3},\\
		\left[\mathcal{W}^2_{m_1},\mathcal{W}^1_{m_2},\mathcal{W}^1_{m_3}\right]&=0.
		\end{align*}
		\item[(ii)]the $(p,q)$-$3$-algebra is derived as:
		\begin{small}
		\begin{align*}
		 \left[\mathcal{W}^2_{m_1},\mathcal{W}^2_{m_2},\mathcal{W}^2_{m_3}\right]&=\frac{1}{4}\big([m_3]_{p,q}-[m_2]_{p,q}\big)\big([m_3]_{p,q}-[m_1]_{p,q}\big)\big([m_2]_{p,q}-[m_1]_{p,q}\big)\,\mathcal{W}^1_{m_1+m_2+m_3},\\
		\left[\mathcal{W}^2_{m_1},\mathcal{W}^2_{m_2},\mathcal{W}^1_{m_3}\right]&=\big([m_2]_{p,q}-[m_1]_{p,q}\big)\,\mathcal{W}^2_{m_1+m_2+m_3},\\
		\left[\mathcal{W}^2_{m_1},\mathcal{W}^1_{m_2},\mathcal{W}^1_{m_3}\right]&=\big([m_2]_{p,q}-[m_3]_{p,q}\big)\,\mathcal{W}^1_{m_1+m_2+m_3},\\
		\left[\mathcal{W}^2_{m_1},\mathcal{W}^1_{m_2},\mathcal{W}^1_{m_3}\right]&=0.
		\end{align*}
			\end{small}
	\end{enumerate}
\end{remark}
Let us consider the $\mathcal{R}(p,q)$-operators
\begin{equation*}
F_m=\mathcal{W}^2_m + \nu\,m\,\mathcal{W}^1_m\quad \mbox{and}\quad R_m=\mathcal{W}^1_m.
\end{equation*}
These operators yield the $\mathcal{R}(p,q)$-Virasoro-Witt algebra
\begin{align*}
\,\left[F_n,\,F_m\right]&=\big([n]_{\mathcal{R}(p,q)}-[m]_{\mathcal{R}(p,q)}\big)\,F_{n+m},\\
\,\left[F_n,\,R_m\right]&=[-m]_{\mathcal{R}(p,q)}\,R_{n+m},\\
\,\left[R_n,\,R_m\right]&=0,
\end{align*}
and the $\mathcal{R}(p,q)$-$3$-algebra
\begin{align}\label{Rpqa14}
\,\left[F_k,\,F_m,\,F_n\right]&=(\frac{1}{4}-\nu^2)\big([n]_{\mathcal{R}(p,q)}-[m]_{\mathcal{R}(p,q)}\big)\big([n]_{\mathcal{R}(p,q)}-[k]_{\mathcal{R}(p,q)}\big)\nonumber\\&\times\big([m]_{\mathcal{R}(p,q)}-[k]_{\mathcal{R}(p,q)}\big)R_{k+m+n},\nonumber\\
\,\left[F_k,\,F_m,\,R_n\right]&=\big([m]_{\mathcal{R}(p,q)}-[k]_{\mathcal{R}(p,q)}\big)\big(F_{k+n+m}-2\nu\,[n]_{\mathcal{R}(p,q)}R_{k+m+n}\big),\nonumber\\
\,\left[F_k,\,R_m,\,R_n\right]&=\big([m]_{\mathcal{R}(p,q)}-[n]_{\mathcal{R}(p,q)}\big)\,R_{k+m+n},\nonumber\\
\,\left[R_k,\,R_m,\,R_n\right]&=0.
\end{align}
By setting
\begin{equation*}
\hat{F}_m=-(\frac{1}{4}-\nu^2)^{-\frac{1}{4}}\,F_m,\quad  \hat{R}_m=(\frac{1}{4}-\nu^2)^{\frac{1}{4}}\,R_m,\quad \mbox{and}\quad z=\frac{2\nu}{(\frac{1}{4}-\nu^2)^{\frac{1}{2}}},
\end{equation*}
where $\nu \geq \pm\frac{1}{2},$ then the $\mathcal{R}(p,q)$-$3$-algebra
\eqref{Rpqa14} becomes
\begin{align}\label{Rpqa15}
\,\left[\hat{F}_k,\,\hat{F}_m,\,\hat{F}_n\right]&=\big([n]_{\mathcal{R}(p,q)}-[m]_{\mathcal{R}(p,q)}\big)\big([n]_{\mathcal{R}(p,q)}-[k]_{\mathcal{R}(p,q)}\big)\nonumber\\&\times\big([k]_{\mathcal{R}(p,q)}-[m]_{\mathcal{R}(p,q)}\big)\hat{R}_{k+m+n},\nonumber\\
\,\left[\hat{F}_k,\,\hat{F}_m,\,\hat{R}_n\right]&=\big([k]_{\mathcal{R}(p,q)}-[m]_{\mathcal{R}(p,q)}\big)\big(\hat{F}_{k+n+m}+z\,[n]_{\mathcal{R}(p,q)}\hat{R}_{k+m+n}\big),\nonumber\\
\,\left[\hat{F}_k,\,\hat{R}_m,\,\hat{R}_n\right]&=\big([n]_{\mathcal{R}(p,q)}-[m]_{\mathcal{R}(p,q)}\big)\,\hat{R}_{k+m+n},\nonumber\\
\,\left[\hat{R}_k,\,\hat{R}_m,\,\hat{R}_n\right]&=0.
\end{align}
We immediately recognize that the $3$-algebra \eqref{Rpqa15} is
nothing but the $\mathcal{R}(p,q)$-Virasoro-Witt $3$-algebra.

For $\mathcal{R}(p,q)=1,$ we derived the Virasoro-Witt $3$-algebra 
introduced
in \cite{CFZ}. When $z=\pm 2i$ with $i=\sqrt{-1}$, it
satisfies the FI \eqref{Rpqa11}.


Now, we study the $\mathcal{R}(p,q)$-$W_{1+\infty}$-$n$-algebra generated by the $\mathcal{R}(p,q)$-operators \eqref{Rpqa1}. We note that the form of $n$-algebra appears to become more complicated. We
present a conjecture as follows:
\begin{align}\label{Rpqa23}
\left[\mathcal{W}^{s_1}_{m_1},\mathcal{W}^{s_2}_{m_2},\cdots,\mathcal{W}^{s_{2n}}_{m_{2n}}\right]&=f(m_1,\cdots, m_{2n},s_1,\cdots,s_{2n})\mathcal{W}^{s_1+\cdots +s_{2n}-(5n-3)}_{m_1+\cdots+m_{2n}}+\cdots,\nonumber\\ \,\left[\mathcal{W}^{s_1}_{m_1},\mathcal{W}^{s_2}_{m_2},\cdots,\mathcal{W}^{s_{2n+1}}_{m_{2n+1}}\right]&=g(m_1,\cdots, m_{2n+1},s_1,\cdots,s_{2n+1})\mathcal{W}^{s_1+\cdots +s_{2n+1}-(5n-2)}_{m_1+\cdots+m_{2n+1}}+\cdots,
\end{align}
where $(\cdots)$ stands for the lower-order terms corresponding
to the quantum effect.

We check the cases of $n$-bracket with $n\leq 7$  and
confirm that the conjecture \eqref{Rpqa23} holds for these cases.
Let us list the several $n$-algebras as follows:
\begin{align*}
\,\left[\mathcal{W}^{s_1}_{m_1},\mathcal{W}^{s_2}_{m_2},\mathcal{W}^{s_{3}}_{m_{3}}\right]&=[m_1(s_3-s_2)+m_2(s_1-s_3)+m_3(s_2-s_1)]_{\mathcal{R}(p,q)}\mathcal{W}^{s_1+s_2+s_3}_{m_1+m_2+m_3}+\cdots,\nonumber\\
\,\left[\mathcal{W}^{s_1}_{m_1},\mathcal{W}^{s_2}_{m_2},\mathcal{W}^{s_{3}}_{m_{3},\mathcal{W}^{s_{4}}_{m_{4}}}\right]&=f(m_1,\cdots, m_{4},s_1,\cdots,s_{4})\mathcal{W}^{s_1+s_2+s_3+s_4-7}_{m_1+\cdots+m_{2n}}+\cdots,\nonumber\\ \,\left[\mathcal{W}^{s_1}_{m_1},\mathcal{W}^{s_2}_{m_2},\cdots,\mathcal{W}^{s_{5}}_{m_{5}}\right]&=g(m_1,\cdots, m_{5},s_1,\cdots,s_{5})\mathcal{W}^{s_1+\cdots +s_{5}-8}_{m_1+\cdots+m_{5}}+\cdots,
\end{align*}
where
\begin{align*}
f(m_1,\cdots, m_{4},s_1,\cdots,s_{4})&=-\sum_{
	A}\bigg(\frac{\mathbb{C}^{r_1}_{s_1-1}\cdots\mathbb{C}^{r_4}_{s_4-1} }{2^{r_1+\cdots+r_4}}\epsilon^{i_1\cdots i_4}_{1\cdots 4}\nonumber\\&\times \mathbb{C}^{\alpha_3}_{s_{i_4}-1-r_{i_4}}\mathbb{C}^{\alpha_2}_{s_{i_3}+s_{i_4}-2-r_{i_4}-r_{i_3}-\alpha_3}\nonumber\\&\times \mathbb{C}^{\alpha_1}_{s_{i_2}+s_{i_3}+s_{i_4}-2-r_{i_4}-r_{i_3}-r_{i_2}-\alpha_3-\alpha_2}\nonumber\\&\times \mathbb{A}^{r_{i_1}+\alpha_1}_{m_{i_1}+s_{i_1}-1}\mathbb{A}^{r_{i_2}+\alpha_2}_{m_{i_2}+s_{i_2}-1}\mathbb{A}^{r_{i_3}+\alpha_3}_{m_{i_3}+s_{i_3}-1}\mathbb{A}^{r_{i_4}+\alpha_4}_{m_{i_4}+s_{i_4}-1}q^{\gamma}p^{\bar{\gamma}} \bigg),
\end{align*}
\begin{align*}
g(m_1,\cdots, m_{5},s_1,\cdots,s_{5})&=-\sum_{
	\bar{A}}\bigg(\frac{\mathbb{C}^{r_1}_{s_1-1}\cdots\mathbb{C}^{r_5}_{s_5-1} }{2^{r_1+\cdots+r_5}}\epsilon^{i_1\cdots i_5}_{1\cdots 5}\mathbb{C}^{\alpha_4}_{s_{i_5}-1-r_{i_5}}\mathbb{C}^{\alpha_3}_{s_{i_4}+s_{i_5}-2-r_{i_4}-r_{i_5}-\alpha_4} \nonumber\\&\times q^{\eta}p^{\bar{\eta}} \,\mathbb{C}^{\alpha_2}_{s_{i_3}+s_{i_4}+s_{i_5}-3-r_{i_4}-r_{i_3}-r_{i_5}-\alpha_3-\alpha_4}\nonumber\\&\times \mathbb{C}^{\alpha_1}_{s_{i_2}+s_{i_3}+s_{i_4}+s_{i_5}-4-r_{i_5}-r_{i_4}-r_{i_3}-r_{i_2}-\alpha_4-\alpha_3-\alpha_2}\nonumber\\&\times \mathbb{A}^{r_{i_1}+\alpha_1}_{m_{i_1}+s_{i_1}-1}\mathbb{A}^{r_{i_2}+\alpha_2}_{m_{i_2}+s_{i_2}-1}\mathbb{A}^{r_{i_3}+\alpha_3}_{m_{i_3}+s_{i_3}-1}\mathbb{A}^{r_{i_4}+\alpha_4}_{m_{i_4}+s_{i_4}-1}\mathbb{A}^{r_{i_4}+\alpha_5}_{m_{i_5}+s_{i_5}-1}\bigg),
\end{align*}
\begin{align*}
A&=\{r_1+\cdots +r_4+\alpha_1+\cdots +\alpha_3=4,\quad r_1,\cdots,r_4\geq 0,\quad \alpha_1,\alpha_2,\alpha_3\geq 0\}\nonumber\\
\bar{A}&=\{r_1+\cdots +r_5+\alpha_1+\cdots +\alpha_4=4,\quad r_1,\cdots,r_5\geq 0,\quad \alpha_1,\cdots,\alpha_4\geq 0\}.\end{align*}

Although it is hard to give the explicit expression
of $\mathcal{R}(p,q)$-$W_{1+\infty}$ $n$-algebra, straightforward calculation gives
the following multibracket for the $\mathcal{R}(p,q)$-operators $\mathcal{W}^s_{m}$
with any fixed index $s\geq 1.$ 
\begin{align}\label{Rpqa28}
\left[\mathcal{W}^{s}_{m_1},\mathcal{W}^{s}_{m_2},\cdots, \mathcal{W}^{s}_{m_{2s-1}}\right]&=\frac{1}{2^{\alpha}}\bigg(\frac{[-2\sum_{l=1}^{n}m_{l}]_{\mathcal{R}(p,q)}}{[-\sum_{l=1}^{n}m_{l}]_{\mathcal{R}(p,q)}}\bigg)^{\alpha}\epsilon_{1 2 \cdots (2s-1)}^{i_1 \cdots i_{2s-1}}
\mathcal{W}_{m_{i_1}}^{s}\cdots 
 \mathcal{W}_{m_{i_{2s-1}}}^{s}\nonumber\\
&=\frac{(-1)^{(2s-1)s}}{2^{\alpha}}\bigg(\frac{[-2\sum_{l=1}^{n}m_{l}]_{\mathcal{R}(p,q)}}{[-\sum_{l=1}^{n}m_{l}]_{\mathcal{R}(p,q)}}\bigg)^{\alpha}\sum_{r_1=0}^{s-1}\cdots\sum_{r_{2s-1}=0}^{s-1}\nonumber\\
&\times \sum_{\alpha_{2s-2
}=0}^{b_{2s-2}}\cdots \sum_{\alpha_1=0}^{b_1}\frac{\prod_{j=1}^{2s-2}\big(\mathbb{C}^{r_j}_{s-1}\mathbb{C}^{\alpha_j}_{b_j}\big)\mathbb{C}^{r_{2s-1}}_{s-1}}{2^{r_1+\cdots+ r_{2s-1}}}
\epsilon_{1 2 \cdots (2s-1)}^{i_1 \cdots i_{2s-1}}q^{\tilde{\gamma}}\,p^{\hat{\gamma}}\nonumber\\&\times \mathbb{A}^{r_1+\alpha_1}_{m_{i_1}+s-1}\cdots \mathbb{A}^{r_{2s-2}+\alpha_{2s-2}}_{m_{i_{2s-2}}+s-1}\mathbb{A}^{r_{2s-1}}_{m_{i_{2s-1}}+s-1}x^{b_0}\partial^{m_1+\cdots+m_{2s-1}+b_0}_{\mathcal{R}(p,q)}\nonumber\\&=\frac{(-1)^{s+1}}{2^{\alpha}}\bigg(\frac{[-2\sum_{l=1}^{n}m_{l}]_{\mathcal{R}(p,q)}}{[-\sum_{l=1}^{n}m_{l}]_{\mathcal{R}(p,q)}}\bigg)^{\alpha}\sum_{r_1=0}^{s-1}\cdots\sum_{r_{2s-1}=0}^{s-1}\nonumber\\&\times \sum_{\alpha_{2s-2
	}=0}^{b_{2s-2}}\cdots \sum_{\alpha_1=0}^{b_1}(-1)^{\sigma}\frac{\prod_{j=1}^{2s-2}\big(\mathbb{C}^{r_j}_{s-1}\mathbb{C}^{\alpha_j}_{b_j}\big)\mathbb{C}^{r_{2s-1}}_{s-1}}{2^{r_1+\cdots+ r_{2s-1}}}\nonumber\\&\times \mathbb{V}_{2s-1}\mathcal{W}^{1}_{m_1+m_2+\cdots+m_{2s-1}},
\end{align}
where $\mathbb{V}_{2s}$ is the Vandermonde determinant 
\begin{equation*}
    \mathbb{V}_{2n}=\prod_{1\leqslant j< k\leqslant 2s} \big( [m_k]_{\mathcal{R}(p,q)}-[m_j]_{\mathcal{R}(p,q)}\big)
\end{equation*}
and 
$b_j=(2s-1-j)(s-1)-r_{2s-1}-\cdots-r_{j+1}-\alpha_{2s-1}-\cdots-\alpha_{j+1},\,j\in\{0,1,\cdots,2s-1\},$ and $\sigma$ denotes the permutation 
\begin{eqnarray*}
	\sigma=\left( \begin{array} {ccccc}
0 &1&\cdots&  2s-3 & 2s-2   \\ 
\alpha_1+r_1 & \alpha_2+r_2& \cdots& \alpha_{2s-2}+r_{2s-2}& r_{2s-1}
\end {array} \right), 
\end{eqnarray*} 
with $\alpha_{2s-1}=0.$

Let us take the scaled $\mathcal{R}(p,q)$-operators $\mathcal{W}^{s}_{m}\longrightarrow Q^{-\frac{1}{2s-1}}_s$, where the deformed scaling coefficient $Q_s$ is given
by the following relation 
\begin{align*}
Q_s&=\frac{(-1)^{s+1}}{2^{\alpha}}\bigg(\frac{[-2\sum_{l=1}^{n}m_{l}]_{\mathcal{R}(p,q)}}{[-\sum_{l=1}^{n}m_{l}]_{\mathcal{R}(p,q)}}\bigg)^{\alpha}\sum_{r_1=0}^{s-1}\cdots\sum_{r_{2s-1}=0}^{s-1}\nonumber\\&\times \sum_{\alpha_{2s-2
	}=0}^{b_{2s-2}}\cdots \sum_{\alpha_1=0}^{b_1}(-1)^{\sigma}\frac{\prod_{j=1}^{2s-2}\big(\mathbb{C}^{r_j}_{s-1}\mathbb{C}^{\alpha_j}_{b_j}\big)\mathbb{C}^{r_{2s-1}}_{s-1}}{2^{r_1+\cdots+ r_{2s-1}}}.
\end{align*}
Then the relation \eqref{Rpqa28} takes the following form:
\begin{equation}\label{Rpqa31}
\left[\mathcal{W}^{s}_{m_1},\mathcal{W}^{s}_{m_2},\cdots, \mathcal{W}^{s}_{m_{2s-1}}\right]=\mathbb{V}_{2s-1}\mathcal{W}^{1}_{m_1+m_2+\cdots+m_{2s-1}}.
\end{equation}
 For $s=2$ in the relation \eqref{Rpqa31}, it gives the $\mathcal{R}(p,q)$-$3$-bracket \eqref{Rpqa9a}. Note that the $\mathcal{R}(p,q)$-operators \eqref{Rpqa1}  do not yield the nontrivial
 $\mathcal{R}(p,q)$-sub-$2s$-algebra \eqref{RpqWnalg}. However, we find that there are
 the closed $\mathcal{R}(p,q)$-$3$-algebra\eqref{Rpqa9a} to \eqref{Rpqa9d} and $\mathcal{R}(p,q)$-$4$-algebra:
 \begin{align}\label{Rpqa32}
 	\,\left[\mathcal{W}^{3}_{m_1},\mathcal{W}^{3}_{m_2},\mathcal{W}^{3}_{m_3},\mathcal{W}^{3}_{m_4}\right]&=-\frac{9}{2}\mathbb{V}_{4}\mathcal{W}^{3}_{m_1+m_2+m_3+m_4}+\frac{1}{8}\mathbb{V}_4\bigg[3 \sum_{1\leq j< k\leq 4}[m_j]_{\mathcal{R}(p,q)}[m_k]_{\mathcal{R}(p,q)}\nonumber\\&+\sum_{j=1}^{4}\big(6[m_j]_{\mathcal{R}(p,q)}+[m_j]^2_{\mathcal{R}(p,q)}\big)+9\bigg]\mathcal{W}^{1}_{m_1+m_2+m_3+m_4}\nonumber\\
 		\,\left[\mathcal{W}^{3}_{m_1},\mathcal{W}^{3}_{m_2},\mathcal{W}^{3}_{m_3},\mathcal{W}^{1}_{m_4}\right]&=-10[m_4]_{\mathcal{R}(p,q)}\mathbb{V}_3\mathcal{W}^{3}_{m_1+m_2+m_3+m_4}+\frac{1}{2}[m_4]_{\mathcal{R}(p,q)}\mathbb{V}_3\nonumber\\&\times\bigg[\sum{1\leq j<k\leq 4}[m_j]_{\mathcal{R}(p,q)}[m_k]_{\mathcal{R}(p,q)}+\big(4-[m_4]_{\mathcal{R}(p,q)}\big)\nonumber\\&\times\big([m_1]_{\mathcal{R}(p,q)}+[m_2]_{\mathcal{R}(p,q)}+[m_3]_{\mathcal{R}(p,q)}\big)+6[m_4]_{\mathcal{R}(p,q)}\nonumber\\&-3[m_4]^2_{\mathcal{R}(p,q)}+4\bigg]\mathcal{W}^{1}_{m_1+m_2+m_3+m_4}\nonumber\\
 		\,\left[\mathcal{W}^{3}_{m_1},\mathcal{W}^{3}_{m_2},\mathcal{W}^{1}_{m_3},\mathcal{W}^{1}_{m_4}\right]&=2[m_3]_{\mathcal{R}(p,q)}[m_4]_{\mathcal{R}(p,q)}\big([m_1]_{\mathcal{R}(p,q)}-[m_2]_{\mathcal{R}(p,q)}\big)\nonumber\\&\times\big([m_3]_{\mathcal{R}(p,q)}-[m_4]_{\mathcal{R}(p,q)}\big)\mathcal{W}^{1}_{m_1+m_2+m_3+m_4}\nonumber\\
 		\,\left[\mathcal{W}^{3}_{m_1},\mathcal{W}^{1}_{m_2},\mathcal{W}^{1}_{m_3},\mathcal{W}^{1}_{m_4}\right]&=0,\nonumber\\
 		\,\left[\mathcal{W}^{1}_{m_1},\mathcal{W}^{1}_{m_2},\mathcal{W}^{1}_{m_3},\mathcal{W}^{1}_{m_4}\right]&=0.
 \end{align}
 Since the $\mathcal{R}(p,q)$-$4$-algebra \eqref{Rpqa32} satisfies the GJI \eqref{ShJ}, it is a $\mathcal{R}(p,q)$-generalized Lie algebra.
 
 From the definition of the  $n$-bracket and using the relation \eqref{Rpqa31}, we have:
 \begin{equation*}
 \left[\mathcal{W}^{s}_{m_1},\mathcal{W}^{s}_{m_2},\cdots,\mathcal{W}^{s}_{m_{2s}}\right]=0.
 \end{equation*}
\section{$\mathcal{R}(p,q)$- multi-variable  $W_{1+\infty}$ algebra and  n-algebra}	We investigate the multi-variable $W_{1+\infty}$ algebra and its n-algebra from the $\mathcal{R}(p,q)$-deformed algebra\cite{HB1}. Relevant particular cases are derived.

Then, from the relation \eqref{r5}, we deduce the  $\mathcal{R}(p,q)$-derivatives $D_j$ in the following form: 
\begin{equation*}
D_j\,f(x_1,\cdots,x_j)=\frac{q-p}{q^{Q}-p^{P}}\mathcal{R}(q^{Q},p^{P})\frac{f(x_1,\cdots,qx_j,\cdots,x_N)-f(x_1,\cdots,px_j,\cdots,x_N) }{(q-p)x_j},
\end{equation*}
where $j\in\{1,2,\cdots,N\}.$
The generalized derivatives $D_j$ satisfy the 
	following  property:
	\begin{equation*}
	\int_{-\infty}^{+\infty}dx_j\,D_j\,f(x_1,\cdots,x_j)=0.
	\end{equation*}

We consider the $\mathcal{R}(p,q)$-operators $\bar{V}^{{\bf r}}_{{\bf m}}$ given by the  relation:
\begin{equation}\label{eq18}
\bar{V}^{{\bf r}}_{{\bf m}}:=\bar{V}^{r_1,\cdots,r_N}_{m_1,\cdots,m_N}=(-1)^{\sum_{j=1}^{N}r_j}\sum_{\sigma \in S_N} \prod_{j=1}^{N} D^{r_{\sigma(j)}-1}_j\,x^{m_{\sigma(j)}+r_{\sigma(j)}-1}_j,
\end{equation}
where $m_j\in\mathbb{Z}$ and $r_j\in\mathbb{Z}_+.$ Then, they obey the commutation relation:
\begin{align}\label{eq21}
[\bar{V}^{{\bf r}}_{{\bf m}}; \bar{V}^{{\bf s}}_{{\bf n}}]&=(-1)^{N}\mathcal{K}(P,Q)\sum_{\sigma \in S_N}\bigg(\sum_{\alpha_1=0}^{s_1-1}\cdots \sum_{\alpha_N=0}^{s_N-1}q^{\mu}\,p^{\tilde{\mu}}\prod_{j=1}^{N}\mathbb{C}^{\alpha_j}_{s_j-1}\mathbb{A}^{\alpha_j}_{m_{\sigma(j)}+r_{\sigma(j)}-1}\nonumber\\&-\sum_{\alpha_1=0}^{r_{\sigma(1)}-1}\cdots \sum_{\alpha_N=0}^{r_{\sigma(N)}-1}q^{\nu}\,p^{\tilde{\nu}}\prod_{j=1}^{N}\mathbb{C}^{\alpha_j}_{r_{\sigma(j)}-1}\mathbb{A}^{\alpha_j}_{n_{j}+s_j-1}\bigg)\bar{V}^{\sigma({\bf r})+{\bf s-\alpha}}_{\sigma({\bf m})+{\bf n}},
\end{align}
where
\begin{eqnarray*}
\left \{
\begin{array}{l}
\mu=\sum_{j=1}^{N}\frac{1}{2}\alpha_j(\alpha_j-1)-(s_j-1)(m_{\sigma(j)}+r_{\sigma(j)}-1)\\
\\
\nu=\sum_{j=1}^{N}\frac{1}{2}\alpha_j(\alpha_j-1)-(r_{\sigma(j)}-1)(n_{j}+s_{j}-1)\\
\\
\tilde{\mu}=\sum_{j=1}^{N}\frac{1}{2}\alpha_j(\alpha_j-1)-j(s_j-1)\\
\\
\tilde{\nu}=\sum_{j=1}^{N}\frac{1}{2}\alpha_j(\alpha_j-1)-j(r_{\sigma(j)}-1).
\end{array}
\right .
\end{eqnarray*}

\begin{remark}
	Taking ${\bf r}={\bf s}=1,$ we obtain the abelian $\mathcal{R}(p,q)$-current algebra.
\end{remark}

Let us consider the $n$-commutator
\begin{equation*}
[\bar{V}^{r_1}_{ m_1},\cdots, \bar{V}^{r_n}_{ m_n}]:=\frac{1}{2^{\alpha}}\bigg(\frac{[-2\sum_{l=1}^{n}m_{l}]_{\mathcal{R}(p,q)}}{[-\sum_{l=1}^{n}m_{l}]_{\mathcal{R}(p,q)}}\bigg)^{\alpha}\epsilon^{i_1\cdots i_n}_{1\cdots n}\,\prod_{j=1}^{n}\bar{V}^{r_{i_j}}_{ m_{i_j}}.
\end{equation*}
By direct calculation of the $n$-commutator \eqref{eq18},
the closed $n$-algebra is given by the relation:
\begin{align}\label{eq23}
[\bar{V}^{r_1}_{ m_1},\cdots, \bar{V}^{r_n}_{ m_n}]&=(-1)^{\frac{1}{2}(n-1)(n+2N)}\,\frac{\mathcal{K}(P,Q)}{2^{\alpha}}\bigg(\frac{[-2\sum_{l=1}^{n}m_{l}]_{\mathcal{R}(p,q)}}{[-\sum_{l=1}^{n}m_{l}]_{\mathcal{R}(p,q)}}\bigg)^{\alpha}\epsilon^{i_1\cdots i_n}_{1\cdots n}\nonumber\\&\times\sum_{\sigma_1\cdots \sigma_{n-1}\in S_N}\sum_{\alpha_{11}=0}^{\beta_{11}}\cdots \sum_{\alpha_{(n-1)N}=0}^{\beta_{(n-1)N}}\prod_{j=1}^{N}\prod_{k=1}^{n-1}\mathbb{C}^{\alpha_{k_j}}_{\beta_{k_j}}\mathbb{A}^{\alpha_{k_j}}_{\gamma_{k_j}}\,q^{\lambda}\,p^{\tilde{\lambda}}\,\bar{V}^{\bar{r}}_{ \bar{m}},
\end{align}
where $\lambda=\frac{1}{2}\displaystyle\sum_{k=1}^{n-1}\sum_{j=1}^{N}\alpha_{k_j}(\alpha_{k_j}-1)-\beta_{k_j}\gamma_{k_j}$ {and $\tilde{\lambda}=\frac{1}{2}\displaystyle\sum_{k=1}^{n-1}\sum_{j=1}^{N}\alpha_{k_j}(\alpha_{k_j}-1)$}

Since the generalized Jacobi identity (GJI) \eqref{ShJ}
is guaranteed to hold, the relation \eqref{eq23} with $n$ even is also a generalized quantum Lie algebra.

Note that, for $\mathcal{R}(p,q)=1,$ the relation \eqref{eq21} is reduced to the generalized $W_{1+\infty}$ algebra  and the relation \eqref{eq23} to the $n$-algebra.

Now we consider the $\mathcal{R}(p,q)$-operators $\bar{W}^r_m$ defined by:
\begin{equation*}
\bar{W}^r_m=\frac{(-1)^{N-1}}{(N-1)!}\bar{V}^{(r,1,\cdots,1)}_{(m,0,\cdots,0)}=(-1)^r\sum_{j=1}^{N}D^{r-1}_j\,x^{m+r-1}_j,
\end{equation*}
where $m\in\mathbb{Z}$ and $ r\in\mathbb{Z}_+.$ It's obtained by using \eqref{eq18}.
Then, from the relation \eqref{eq21},  we derive 
the $\mathcal{R}(p,q)$-$W_{1+\infty}$  by the commutation relation:
\begin{align}\label{eq22}
[\bar{W}^r_m;\bar{W}^s_n]=
\mathcal{K}(P,Q)
\bigg(\sum_{k=0}^{r-1}q^{\bar{\mu}}\,p^{\bar{\gamma}}\mathbb{C}^{k}_{r-1}\mathbb{A}^{k}_{n+s-1}-\sum_{k=0}^{s-1}q^{\bar{\nu}}\,p^{\tilde{\gamma}}\mathbb{C}^{k}_{s-1}\mathbb{A}^{k}_{m+r-1}\bigg)\bar{W}^{r+s-1-k}_{m+n},
\end{align}
where $\bar{\mu}=\frac{1}{2}k(k-1)-(r-1)(n+s-1)$ and $\bar{\nu}=\frac{1}{2}k(k-1)-(s-1)(m+r-1).$

\section{ $\mathcal{R}(p,q)$-$W_{\infty}$ constraints for the elliptic hermitian matrix model}
In this section, we investigate the $W_{\infty}$ constraints for the elliptic hermitian matrix model from the generalized quantum algebra introduced in \cite{HB1}. We give the first $\mathcal{R}(p,q)$-operators and derive particular cases related from quantum algebras.

The generalized quantum theta function  $\Gamma(x; \mathcal{R}(p,q))$ is defined  by the relation:
\begin{equation*}\label{Rpqtf}
\Gamma(x; \mathcal{R}(p,q))=\prod\limits_{k=0}^{\infty}\bigg( 1-F\big(\frac{q^{k}}{ p^k}\,x\big)G(P,Q)\bigg)\prod\limits_{k=0}^{\infty}\bigg( 1-F\big(\frac{q^{k+1}}{ p^{k+1}}\,x^{-1}\big)G(P,Q)\bigg),
\end{equation*}
where the functions $F$ and $G$ satisfy the following relation:
\begin{align*}
\,F(x)&=x\nonumber\\
\,G(P,Q)&=\frac{q^{Q}-p^{P}}{q^{Q}\mathcal{R}(p^P,q^Q)},\quad \mbox{if}\quad \mathcal{R}(1,0)=0.
\end{align*}
and 
\begin{align*}
F(x)&= \frac{x}{x-\mathcal{R}(1,0)}, \nonumber
\\
G(P,Q)&= \frac{p(Q-P)\mathcal{R}(pP,qQ)+(pP-qQ)\mathcal{R}(1,0)}{qQ\mathcal{R}(pP,qQ)}, \quad \mbox{if}\quad \mathcal{R}(1,0)\neq 0.
\end{align*}
Moreover,  the generalized generating function for the elliptic hermitian matrix model from the quantum algebras is defined as follows:
\begin{equation}\label{eq17}
Z^{ell}_{N}(\{t\})=\oint\prod_{i=1}^{N}\frac{dx_i}{x_i}\prod_{i<j}\Gamma\big(\frac{x_i}{x_j}; \mathcal{R}(p,q)\big)\Gamma\big(\frac{x_j}{x_i}; \mathcal{R}(p,q)\big)\,\mathrm{Exp}\bigg(\sum_{k=0}^{\infty}\frac{t_k}{k!}\sum_{i=1}^{N}x^k_i\bigg),
\end{equation}
where $\{t\}=\{t_k|k\in\mathbb{N}\}$ and the integration is over the contour around the origin.

The relation \eqref{eq17} may be viewed as the generalized generating function for the expectation values of supersymmetric Wilson loop in different representations.

Let $\widetilde{W}^r_{m}$ be the generalized quantum differential operators with respect to $\{t\}$ such that they satisfy
\begin{equation}\label{eq24}
\widetilde{W}^r_{m}Z^{ell}_{N}(\{t\})=-\oint\prod_{i=1}^{N}\frac{dx_i}{x_i}\bar{W}^r_{m}\prod_{i<j}\Gamma\big(\frac{x_i}{x_j}; \mathcal{R}(p,q)\big)\Gamma\big(\frac{x_j}{x_i}; \mathcal{R}(p,q)\big)\mathrm{exp}\bigg(\sum_{k=0}^{\infty}\frac{t_k}{k!}\sum_{i=1}^{N}x^k_i\bigg),
\end{equation}
with $r>2.$
Then the following constraints can be  obtained from the relation \eqref{eq24}
\begin{equation}\label{eq25}
\widetilde{W}^r_{m}Z^{ell}_{N}(\{t\})=0,
\end{equation}
with
\begin{align}\label{Rpq26}
\widetilde{W}^r_{m}&=\bigg(\frac{K(P,Q)}{q-p}\bigg)^{r-1}\,p^{m+r-1}\,\bigg(\sum_{j=0}^{r-1}(-1)^j\,\mathbb{C}^{j}_{r-1}\,\big(\frac{q}{p}\big)^{\bar{\alpha}}\, \sum_{l=0}^{\infty}\frac{(l+m-2(r-1-j)N)!}{l!}\nonumber\\&\times B_l\big(t^{r-j-1}_1,\cdots,t^{r-j-1}_l\big)\mathcal{D}^{2(r-j-1)}_N\frac{\partial}{\partial_{t_{l+m-2(r-1-j)N}}}\bigg),
\end{align}
where $\bar{\alpha}=(r-j-1)m-(r-j-1)^2N+\frac{1}{2}(r-j-1)(3r-3j-4),$  $B_k$ are the coefficients of the expansion 
\begin{equation}\label{RpqB}
\mathrm{exp}\bigg(\sum_{s=0}^{\infty}\frac{t_s}{s!}x^s\bigg)=\sum_{k=0}^{\infty}\frac{1}{k!}B_k\big(t_1,\cdots,t_k\big)x^k,
\end{equation}
with $t^a_k=\big(q^{ak}-p^{ak}\big)t_k,$ and  $\mathcal{D}^m_{N}$ are the operators given by the relation:
\begin{eqnarray}\label{diffop}
\mathcal{D}^m_N =\frac{1}{N!}\left|\begin{array}{ccccc}
m! \frac{\partial}{\partial t_m} & 1 & 0  &\dots  &0  \\
(2m)!\frac{\partial}{\partial t_{2m}}  &m! \frac{\partial}{\partial t_m} & 2 & \dots & 0\\
\vdots   &  \vdots  &\vdots   &\ddots&\vdots\\  
(m(N-1))! \frac{\partial}{\partial t_{m(N-1)}} & (m(N-2))!\frac{\partial}{\partial t_{m(N-2)}} & \dots  & m!\frac{\partial}{\partial t_m} & N-1\\
(mN)! \frac{\partial}{\partial t_{mN}} & (m(N-1))!\frac{\partial }{\partial t_{m(N-1)}} & \dots  & (2m)!\frac{\partial}{\partial t_{2m}} & m! \frac{\partial}{\partial t_m} 
\end{array}\right|~,
\end{eqnarray}
with the property 
\begin{equation*}
	\mathcal{D}^m_N  \left[ \mathrm{exp}\bigg(\sum\limits_{k=0}^{\infty} \frac{t_k}{k!}\sum\limits_{i=1}^N x_i^k\bigg) \right]=\prod_{j=1}^{N} x_j^m ~ \mathrm{exp}\bigg(\sum\limits_{k=0}^{\infty} \frac{t_k}{k!}\sum\limits_{i=1}^N x_i^k\bigg).
\end{equation*}	
\begin{remark}
	The first several ones of \eqref{Rpq26} can be deduced by: \begin{enumerate}
		\item[(i)] For $r=2,$ we have
		\begin{align}\label{28a}
		\widetilde{W}^2_{m}&=-\frac{K(P,Q)}{q-p}\,p^{m+1}\bigg(\big(\frac{q}{p}\big)^{m-N+1} \sum_{k=0}^{\infty}\frac{(k+m-2N)!}{k!}B_k\big(t^{1}_1,\cdots,t^{1}_k\big)\mathcal{D}^{2}_N\frac{\partial}{\partial_{t_{k+m-2N}}}\nonumber\\&-m!\frac{\partial}{\partial_{t_m}}\bigg).
		\end{align}
		\item[(ii)] For $r=3,$ 
		\begin{align*}
		\widetilde{W}^3_{m}&=\frac{K(P,Q)}{(q-p)^2}\,p^{m+2}\bigg(\big(\frac{q}{p}\big)^{2m-4N+5} \sum_{k=0}^{\infty}\frac{(k+m-4N)!}{k!}B_k\big(t^{2}_1,\cdots,t^{2}_k\big)\mathcal{D}^{4}_N\frac{\partial}{\partial_{t_{k+m-4N}}}\nonumber\\& -\big(\frac{q}{p}\big)^{m+1-N}[2]_{\mathcal{R}(p,q)}\sum_{k=0}^{\infty}\frac{(k+m-2N)!}{k!}B_k\big(t^{1}_1,\cdots,t^{1}_k\big)\mathcal{D}^{2}_N\frac{\partial}{\partial_{t_{k+m-2N}}}+m!\frac{\partial}{\partial_{t_m}}\bigg).
		\end{align*}
		\item[(ii)] For $r=4,$ 
		\begin{align*}
		\widetilde{W}^4_{m}&=-\frac{K(P,Q)}{(q-p)^3}\,p^{m+3}\bigg(\big(\frac{q}{p}\big)^{3m-9N+12} \sum_{k=0}^{\infty}\frac{(k+m-6N)!}{k!}B_k\big(t^{3}_1,\cdots,t^{3}_k\big)\mathcal{D}^{6}_N\frac{\partial}{\partial_{t_{k+m-6N}}}\nonumber\\&-[3]_{\mathcal{R}(p,q)}\big(\frac{q}{p}\big)^{2m-4N+5} \sum_{k=0}^{\infty}\frac{(k+m-4N)!}{k!}B_k\big(t^{2}_1,\cdots,t^{2}_k\big)\mathcal{D}^{4}_N\frac{\partial}{\partial_{t_{k+m-4N}}}\nonumber\\& +[3]_{\mathcal{R}(p,q)}\big(\frac{q}{p}\big)^{m+1-N}\sum_{k=0}^{\infty}\frac{(k+m-2N)!}{k!}B_k\big(t^{1}_1,\cdots,t^{1}_k\big)\mathcal{D}^{2}_N\frac{\partial}{\partial_{t_{k+m-2N}}}-m!\frac{\partial}{\partial_{t_m}}\bigg).
		\end{align*}
	\end{enumerate}
\end{remark}
Note that the operators 
$\widetilde{W}^2_{m}$ \eqref{28a} are the so-called $\mathcal{R}(p,q)$-Virasoro operators where the $q$-Virasoro operarors can be found in \cite{NZ}. The simple computation shows that the operators 
$\widetilde{W}^r_{m}$ \eqref{Rpq26} including the $\mathcal{R}(p,q)$-Virasoro operators do not yield the closed algebra. Let us now apply the same strategy of carrying out the action of the operators on the partition function as done in \cite{NZ}. Then we obtain that the commutation relation of
$\widetilde{W}^r_{m}$ is isomorphic to \eqref{eq22} which does not contain the abelian current algebra. Thus we call \eqref{eq25} the $\mathcal{R}(p,q)$-$W_{\infty}$ constraints.

Although we derive the desired $\mathcal{R}(p,q)$-$W_{\infty}$ algebra in this way, we note that the $n$-bracket of the $\mathcal{R}(p,q)$-$W_{\infty}$ operators \eqref{Rpq26} still does not form a closed $n$-algebra.
We introduce the operators $\widetilde{V}^{{\bf r}}_{{\bf m}}$
which satisfy
\begin{equation}\label{eq29}
\widetilde{V}^{{\bf r}}_{{\bf m}}Z^{ell}_{N}(\{t\})=-\oint\prod_{i=1}^{N}\frac{dx_i}{x_i}\bar{V}^{{\bf r}}_{{\bf m}}\prod_{i<j}\Gamma\big(\frac{x_i}{x_j}; \mathcal{R}(p,q)\big)\Gamma\big(\frac{x_j}{x_i}; \mathcal{R}(p,q)\big) \mathrm{exp}\bigg(\sum_{k=0}^{\infty}\frac{t_k}{k!}\sum_{i=1}^{N}x^k_i\bigg),
\end{equation}
where ${\bf r}\neq (1,1,\cdots,1).$
It is obvious that the right-hand side of \eqref{eq29} equals zero. Thus we have
\begin{equation*}\label{eq30}
\widetilde{V}^{{\bf r}}_{{\bf m}}Z^{ell}_{N}(\{t\})=0.
\end{equation*}
Substituting $\bar{V}^{{\bf r}}_{{\bf m}}$ into \eqref{eq29}, we may derive the operators $\widetilde{V}^{{\bf r}}_{{\bf m}}$ as follows:
\begin{align}\label{eq31}
\widetilde{V}^{{\bf r}}_{{\bf m}}&=\bigg(-\frac{K(P,Q)}{q-p}\bigg)^{\sum_{i=1}^{N}r_i-N}\,p^{m_{\alpha}+r_{\alpha}-1}\bigg(\sum_{j_1=0}^{r_1-1}\cdots\sum_{j_N=0}^{r_N-1} \sum_{k_1=0}^{\infty}\cdots \sum_{k_N=0}^{\infty}\frac{(-1)^{j_1+\vdots+j_N}}{k_1!\cdots k_N!}\nonumber\\&\times\prod_{i=1}^{N}\mathbb{C}^{j_i}_{r_i-1}\big(\frac{q}{p}\big)^{\rho} B_{k_1}\big(t^{r_1-j_1-1}_1,\cdots,t^{r_1-j_1-1}_{k_1}\big)\cdots B_{k_N}\big(t^{r_N-j_N-1}_1,\cdots,t^{r_N-j_N-1}_{k_N}\big)\nonumber\\&\times\mathcal{D}^{2\sum_{\alpha=1}^{N}(r_{\alpha}-1-j_{\alpha})}_N \mathcal{D}_{k_1+m_1-2(r_1-j_1-1)N,\cdots,k_N+m_N-2(r_N-j_N-1)N}\bigg),
\end{align}
where
\begin{align*}
\rho&=\sum_{\alpha=1}^{N}(r_{\alpha}-1-j_{\alpha})m_{\alpha}-(r_{\alpha}-1-j_{\alpha})^{2}N+\frac{1}{2}(r_{\alpha}-1-j_{\alpha})(3r_{\alpha}-4-3j_{\alpha})\nonumber\\&+2\sum_{\alpha\beta}(r_{\beta}-1-j_{\beta})(r_{\alpha}-1-j_{\alpha})
\end{align*}
and the operators  $\mathcal{D}_{m_1,m_2,\cdots,m_N}$ satisfy
\begin{eqnarray}\label{eq32}
\mathcal{D}_{m_1,m_2,\cdots,m_N} \left[\mathrm{exp}\bigg(\sum\limits_{k=0}^{\infty} \frac{t_k}{k!}\sum\limits_{i=1}^N x_i^k \bigg) \right]=\bigg(\sum_{\sigma \in S_n}\prod_{j=1}^{N} x_j^{m_{\sigma(j)}}\bigg) ~ ~ \mathrm{exp}\bigg(\sum\limits_{k=0}^{\infty} \frac{t_k}{k!}\sum\limits_{i=1}^N x_i^k\bigg).
\end{eqnarray}
From the relation \eqref{eq32}, it is obvious to have $\mathcal{D}_{m}=m!\,\frac{\partial}{\partial_{t_m}}.$
A series of $\mathcal{D}_{m_1,m_2,\cdots,m_N}$ can be introduced recursively as
\begin{equation*}
\mathcal{D}_{m_1,m_2,\cdots,m_N}=\frac{1}{N}\sum_{k=1}^{N}(-1)^{k-1}k!\,\sum_{S\in M_{n,k}}(s_1+\cdots+s_k)!\,\frac{\partial}{\partial_{t_{s_1+\cdots+s_k}}}\,\mathcal{D}_{m_1,\cdots,\hat{s}_1,\cdots,\hat{s}_k,\cdots,m_N},
\end{equation*}
where $M_N=\{m_1,\cdots,m_N\}$, $M_{N,k}$ is the collection of all subsets of $M_N$ with $k$ elements, $S=\{s_1,s_2,\cdots,s_k\}$ stands for the term that is omitted. It can be easily seen from \eqref{eq31} $\widetilde{W}^{r}_{ m}=\widetilde{V}^{ (r,1,\cdots,1)}_{ (m,0,\cdots,0)}.$

After a straightforward computation, we obtain the product relation
\begin{align}\label{eq34}
\widetilde{V}^{{\bf r}}_{{\bf m}}\widetilde{V}^{{\bf s}}_{{\bf n}}&\sim \sum_{\sigma\in S_N}\sum_{\alpha_1=0}^{r_1-1}\cdots\sum_{\alpha_N=0}^{r_N-1} q^{\sum_{j=1}^{N}\frac{1}{2}\alpha_j(\alpha_j-1)-(r_j-1)(n_{\sigma(j)}+s_{\sigma(j)}-1)}\nonumber\\&\times \bigg(\prod_{j=1}^{N}\mathbb{C}^{\alpha_j}_{r_j-1}\,\mathbb{A}^{\alpha_j}_{n_{\sigma(j)}+s_{\sigma(j)}-1}\bigg)\widetilde{V}^{{\bf r}+\sigma({\bf s})-{\bf \alpha}}_{{\bf m}+\sigma({\bf n})},
\end{align}
where $\sim$ means that the right and left hand sides are equivalent upon the action on the partition function.

Based on the equivalent relation \eqref{eq34}, it is easy to see that the commutation relation of 
$\widetilde{V}^{{\bf r}}_{{\bf m}}$ is isomorphic to \eqref{eq21} without the abelian current algebra. Hence we call \eqref{eq30} the generalized $\mathcal{R}(p,q)$-$W_{\infty}$
constraints. In addition to the infinite-dimensional Lie algebra, it is not hard to prove that $n$-commutator of the operators 
$\widetilde{V}^{{\bf r}}_{{\bf m}}$ also yields the closed n-algebra, that is isomorphic to \eqref{eq23} in which the abelian current operators are truncated. Thus we uncover the higher algebraic structures in the elliptic matrix model in this way.
\begin{remark} Here, we  list the $W_{\infty}$ constraints for the elliptic hermitian matrix model associated with some qauntum algebras presented in the literature. 
	\begin{enumerate}
		\item[(i)] For $\mathcal{R}(x)=\frac{x-1}{q-1},$ we recovered the $q$-$W_{\infty}$ constraints for the elliptic hermitian matrix model given in \cite{WWYZZ}.
		\item[(ii)] Taking $\mathcal{R}(u,v)=(p-q)^{-1}(u-v),$ we obtain the results 
		corresponding to the $(p,q)$-deformed algebra:
		The $(p,q)$-theta function  $\Gamma(x; p,q)$ is given:
		\begin{equation*}
		\Gamma(x; p,q)=\prod\limits_{k=0}^{\infty}\big( p^k-q^{k}\,x\big)\prod\limits_{k=0}^{\infty}\bigg( p^{k+1}-q^{k+1}.x^{-1}\big),
		\end{equation*}
		
		Also,  the $(p,q)$-generating function for the elliptic hermitian matrix model is deduced by:
		\begin{equation*}\label{pq17}
		Z^{ell}_{N}(\{t\})=\oint\prod_{i=1}^{N}\frac{dx_i}{x_i}\prod_{i<j}\Gamma\big(\frac{x_i}{x_j}; p,q\big)\Gamma\big(\frac{x_j}{x_i}; p,q\big)\,\mathrm{exp}\bigg(\sum_{k=0}^{\infty}\frac{t_k}{k!}\sum_{i=1}^{N}x^k_i\bigg),
		\end{equation*}
		where $\{t\}=\{t_k|k\in\mathbb{N}\}.$ 
		%
		
		We consider the operators $\widetilde{W}^r_{m}$ satisfying
		\begin{equation*}
		\widetilde{W}^r_{m}Z^{ell}_{N}(\{t\})=-\oint\prod_{i=1}^{N}\frac{dx_i}{x_i}\bar{W}^r_{m}\prod_{i<j}\Gamma\big(\frac{x_i}{x_j}; p,q\big)\Gamma\big(\frac{x_j}{x_i}; p,q\big)\mathrm{exp}\bigg(\sum_{k=0}^{\infty}\frac{t_k}{k!}\sum_{i=1}^{N}x^k_i\bigg),\quad r>2.
		\end{equation*}
		Then from the  constraints \eqref{eq25},  
		we have:
		\begin{align*}
		\widetilde{W}^r_{m}&=\bigg(\frac{1}{q-p}\bigg)^{r-1}\,p^{m+r-1}\,\bigg(\sum_{j=0}^{r-1}(-1)^j\,\mathbb{C}^{j}_{r-1}\,\big(\frac{q}{p}\big)^{\bar{\alpha}}\, \sum_{l=0}^{\infty}\frac{(l+m-2(r-1-j)N)!}{l!}\\&\times B_l\big(t^{r-j-1}_1,\cdots,t^{r-j-1}_l\big)\mathcal{D}^{2(r-j-1)}_N\frac{\partial}{\partial_{t_{l+m-2(r-1-j)N}}}\bigg),
		\end{align*}
		where $\bar{\alpha}=(r-j-1)m-(r-j-1)^2N+\frac{1}{2}(r-j-1)(3r-3j-4),$  $B_k$ and $\mathcal{D}^m_{N}$ are given by \eqref{RpqB} and \eqref{diffop}, respectively.
		\begin{remark}
			The first $(p,q)$-operators are deduced by: \begin{enumerate}
				\item[(i)] For $r=2,$ we have
				\begin{align*}
				\widetilde{W}^2_{m}&=-\frac{1}{q-p}\,p^{m+1}\bigg(\big(\frac{q}{p}\big)^{m-N+1} \sum_{k=0}^{\infty}\frac{(k+m-2N)!}{k!}B_k\big(t^{1}_1,\cdots,t^{1}_k\big)\mathcal{D}^{2}_N\frac{\partial}{\partial_{t_{k+m-2N}}}\\&-m!\frac{\partial}{\partial_{t_m}}\bigg).
				\end{align*}
				\item[(ii)] For $r=3,$ 
				\begin{align*}
				\widetilde{W}^3_{m}&=\frac{1}{(q-p)^2}\,p^{m+2}\bigg(\big(\frac{q}{p}\big)^{2m-4N+5} \sum_{k=0}^{\infty}\frac{(k+m-4N)!}{k!}B_k\big(t^{2}_1,\cdots,t^{2}_k\big)\mathcal{D}^{4}_N\frac{\partial}{\partial_{t_{k+m-4N}}}\\& -\big(\frac{q}{p}\big)^{m+1-N}[2]_{\mathcal{R}(p,q)}\sum_{k=0}^{\infty}\frac{(k+m-2N)!}{k!}B_k\big(t^{1}_1,\cdots,t^{1}_k\big)\mathcal{D}^{2}_N\frac{\partial}{\partial_{t_{k+m-2N}}}+m!\frac{\partial}{\partial_{t_m}}\bigg).
				\end{align*}
				\item[(ii)] For $r=4,$ 
				\begin{align*}
				\widetilde{W}^4_{m}&=-\frac{1}{(q-p)^3}\,p^{m+3}\bigg(\big(\frac{q}{p}\big)^{3m-9N+12} \sum_{k=0}^{\infty}\frac{(k+m-6N)!}{k!}B_k\big(t^{3}_1,\cdots,t^{3}_k\big)\mathcal{D}^{6}_N\frac{\partial}{\partial_{t_{k+m-6N}}}\nonumber\\&-[3]_{\mathcal{R}(p,q)}\big(\frac{q}{p}\big)^{2m-4N+5} \sum_{k=0}^{\infty}\frac{(k+m-4N)!}{k!}B_k\big(t^{2}_1,\cdots,t^{2}_k\big)\mathcal{D}^{4}_N\frac{\partial}{\partial_{t_{k+m-4N}}}\nonumber\\& +[3]_{\mathcal{R}(p,q)}\big(\frac{q}{p}\big)^{m+1-N}\sum_{k=0}^{\infty}\frac{(k+m-2N)!}{k!}B_k\big(t^{1}_1,\cdots,t^{1}_k\big)\mathcal{D}^{2}_N\frac{\partial}{\partial_{t_{k+m-2N}}}-m!\frac{\partial}{\partial_{t_m}}\bigg).
				\end{align*}
			\end{enumerate}
		\end{remark}
		
		
		We introduce the $(p,q)$-operators $\widetilde{V}^{{\bf r}}_{{\bf m}}$
		which obey
		\begin{equation*}
		\widetilde{V}^{{\bf r}}_{{\bf m}}Z^{ell}_{N}(\{t\})=-\oint\prod_{i=1}^{N}\frac{dx_i}{x_i}\bar{V}^{{\bf r}}_{{\bf m}}\prod_{i<j}\Gamma\big(\frac{x_i}{x_j}; p,q\big)\Gamma\big(\frac{x_j}{x_i}; p,q\big) \mathrm{exp}\bigg(\sum_{k=0}^{\infty}\frac{t_k}{k!}\sum_{i=1}^{N}x^k_i\bigg),
		\end{equation*}
		where ${\bf r}\neq (1,1,\cdots,1).$
		It is obvious that the right-hand side of \eqref{eq29} equals zero. Thus we have
		\begin{equation*}
		\widetilde{V}^{{\bf r}}_{{\bf m}}Z^{ell}_{N}(\{t\})=0.
		\end{equation*}
		Substituting $\bar{V}^{{\bf r}}_{{\bf m}}$ into \eqref{eq29}, we may derive the operators $\widetilde{V}^{{\bf r}}_{{\bf m}}$ as follows:
		\begin{align*}
		\widetilde{V}^{{\bf r}}_{{\bf m}}&=\bigg(-\frac{1}{q-p}\bigg)^{\sum_{i=1}^{N}r_i-N}\,p^{m_{\alpha}+r_{\alpha}-1}\bigg(\sum_{j_1=0}^{r_1-1}\cdots\sum_{j_N=0}^{r_N-1} \sum_{k_1=0}^{\infty}\cdots \sum_{k_N=0}^{\infty}\frac{(-1)^{j_1+\vdots+j_N}}{k_1!\cdots k_N!}\nonumber\\&\times\prod_{i=1}^{N}\mathbb{C}^{j_i}_{r_i-1}\big(\frac{q}{p}\big)^{\rho} B_{k_1}\big(t^{r_1-j_1-1}_1,\cdots,t^{r_1-j_1-1}_{k_1}\big)\cdots B_{k_N}\big(t^{r_N-j_N-1}_1,\cdots,t^{r_N-j_N-1}_{k_N}\big)\nonumber\\&\times\mathcal{D}^{2\sum_{\alpha=1}^{N}(r_{\alpha}-1-j_{\alpha})}_N \mathcal{D}_{k_1+m_1-2(r_1-j_1-1)N,\cdots,k_N+m_N-2(r_N-j_N-1)N}\bigg),
		\end{align*}
		where
		\begin{align*}
		\rho&=\sum_{\alpha=1}^{N}(r_{\alpha}-1-j_{\alpha})m_{\alpha}-(r_{\alpha}-1-j_{\alpha})^{2}N+\frac{1}{2}(r_{\alpha}-1-j_{\alpha})(3r_{\alpha}-4-3j_{\alpha})\nonumber\\&+2\sum_{\alpha\beta}(r_{\beta}-1-j_{\beta})(r_{\alpha}-1-j_{\alpha}).
		\end{align*} 
		\item[(iii)] It is worth noticing that one can also consider the meromorphic
		function 
		\begin{equation*}
		\mathcal{R}(x,y)=\frac{x-y}{a\frac{x}{p}-b\frac{y}{q}
		}
		\end{equation*}
		where $a,$ $b$ are complex numbers
		and deduce the above results.
	\end{enumerate}
\end{remark}
\section{ $\mathcal{R}(p,q)$-$W_{\infty}$ constraints for a toy model} In this section, we investigate the $W_{\infty}$ constraints for a toy model induced by the quantum algebra \cite{HB1}. Moreover, we deduce particular cases.

 Let us take the $\mathcal{R}(p,q)$-operators
	\begin{equation}\label{aRpqa}
	_aW^r_m=-\mathcal{D}^{r-1}_{\mathcal{R}(p^{a},q^{a})}\,x^{m+r-1},\quad m \in\mathbb{Z},\quad r\in\mathbb{Z}_{+},
	\end{equation}
	where  $\mathcal{D}_{\mathcal{R}(p^{a},q^{a})}$ is the deformed derivative given by:
	\begin{equation*}
	\mathcal{D}_{\mathcal{R}(p^{a},q^{a})}f(x)=\frac{p-q}{p^{P}-q^{Q}}\mathcal{R}(p^{P},q^{Q})\frac{f(p^a\,x)-f(q^a\,x)}{(p^a-q^a)x},\quad a>1.
	\end{equation*}
	The relation \eqref{aRpqa} can be rewritten as follows:
	\begin{align*}
	_aW^r_m&=-\, _{a}\mathbb{A}^{r-1}_{m+r-1}\,x^{m},\nonumber\\
	&=\prod_{i=1}^{r-1}\,T^{\mathcal{R}(p^{a},q^{a})}_{m+r-i-1},
	\end{align*}
	where $T^{\mathcal{R}(p^{a},q^{a})}_{m}=-\mathcal{D}_{\mathcal{R}(p^{a},q^{a})}\,x^{m+1}$ is the $\mathcal{R}(p,q)$-operators introduced in \cite{HMM}. 
	
	
	Then, we can deduced that the $\mathcal{R}(p,q)$-operators $_aW^{r}_m$ is the generalization of the $\mathcal{R}(p,q)$-operators 
	$T^{\mathcal{R}(p^{a},q^{a})}_m$.

We consider the generating function given by 
	presented by  \cite{NZ}: \begin{equation}
	Z^{toy}(t)=\int \, \,x^{\gamma}\,\mathrm{exp}\bigg(\displaystyle\sum_{s=0}^{\infty}{t_s\over s!}x^s\bigg)\,dx.\end{equation}
	We suppose that the $\mathcal{R}(p,q)$-derivative $\mathcal{D}^{r}_{\mathcal{R}(p^{a},q^{a})}$ has the property:
	\begin{equation*}
	\int_{\mathbb{R}}\,\mathcal{D}^{r-1}_{\mathcal{R}(p^{a},q^{a})} \,f(x)d\,x=0, \quad r> 1.
	\end{equation*} 
	Then,  we have
	\begin{equation}\label{qt4}
	\int_{-\infty}^{+\infty}\mathcal{D}^{r-1}_{\mathcal{R}(p^{a},q^{a})}\,\left(x^{m+r+\gamma-1}\,\mathrm{exp}\bigg(\sum_{s=0}^{\infty}{t_s\over s!}x^s\bigg)\right)d\,x=0.
	\end{equation}
	For the exponent in the relation \eqref{qt4}, we consider the expression given by the relation \eqref{RpqB}.
	 Then, we can rewritte the integrand in the relation\eqref{qt4} as follows:
	\begin{align*}\label{qt6}
	&\mathcal{D}^{r-1}_{\mathcal{R}(p^{a},q^{a})} \bigg(x^{m+\gamma+r-1}\,\mathrm{exp}\bigg(\displaystyle\sum_{s=0}^{\infty}{t_s\over s!}x^s\bigg)\bigg)
	=\bigg(\sum_{i=0}^{r-1}\mathbb{C}^{i}_{r-1}\,_a\mathbb{A}^{i}_{m+r+\gamma-1}q^{a(r-1-i)(m+r+\gamma-1)}\nonumber\\ &\qquad\qquad\qquad\times \sum_{k=1}^{\infty}p^{a\,i\,k}\frac{B_{(r-1-i)k}(t_1,\cdots,t_k)}{k(r-1-i)!}\,_a\mathbb{A}^{r-1-i}_{k}\,x^{m+k(r-1-i)}\bigg)
x^{\gamma}	\mathrm{exp}\bigg(\displaystyle\sum_{s=0}^{\infty}{t_s\over s!}x^s\bigg).
	\end{align*}
	 Then,  the identity \eqref{qt4} implies the constraints on the partition function\cite{NZ}: 
	\begin{equation}\label{qt7}
	_a\bar{W}^r_m\,Z^{(toy)}(t)=0,\quad m\geq 0,\end{equation}
	where
	\begin{align}
	_a\bar{W}^r_m&=\sum_{i=0}^{r-1}\mathbb{C}^{i}_{r-1}\,_a\mathbb{A}^{i}_{m+r+\gamma-1}q^{a(r-1-i)(m+r+\gamma-1)}\nonumber\\&\times \sum_{k=1}^{\infty}p^{aik}\frac{(m+k(r-1-i))!}{k!}B_{k(r-1-i)}(t_1,\cdots,t_k)\,_a\mathbb{A}^{r-1-i}_{k}\frac{\partial}{\partial t_{m+k(r-1-i)}}.
	\end{align}
\begin{example}
	 We derive the first $\mathcal{R}(p,q)$-operators $_a\bar{W}^r_m$ as follows:
\begin{enumerate}
	\item[(a)] Taking $r=2,$ we have:
	\begin{align*}
	_a\bar{W}^2_m&=q^{a(m+1+\gamma)}\sum_{k=1}^{\infty}\frac{(m+k)!}{k!}B_{k}(t_1,\cdots,t_k)\,_a\mathbb{A}^{1}_{k}\frac{\partial}{\partial t_{m+k}}\nonumber\\&+[m+\gamma+1]_{\mathcal{R}(p,q)}m!\frac{\partial}{\partial t_{m}}\sum_{k=1}^{\infty}p^{ak}.
	\end{align*}
	\item[(b)] For $r=3,$ we get:
	\begin{align*}
	_a\bar{W}^3_m&=\sum_{i=0}^{2}\mathbb{C}^{i}_{2}\,_a\mathbb{A}^{i}_{m+2+\gamma}q^{a(2-i)(m+r+\gamma-1)}\nonumber\\&\times \sum_{k=1}^{\infty}p^{aik}\frac{(m+k(2-i))!}{k(2-i)!}B_k(t_1,\cdots,t_k)\,_a\mathbb{A}^{2-i}_{k}\frac{\partial}{\partial t_{m+k(2-i)}}.
	\end{align*}
	\item[(c)] Setting $r=4,$ we obtain:
		\begin{align*}
	_a\bar{W}^4_m&=\sum_{i=0}^{3}\mathbb{C}^{i}_{3}\,_a\mathbb{A}^{i}_{m+3+\gamma}q^{a(3-i)(m+3+\gamma)}\nonumber\\&\times \sum_{k=1}^{\infty}p^{aik}\frac{(m+k(3-i))!}{k(3-i)!}B_{k(3-i)}(t_1,\cdots,t_k)\,_a\mathbb{A}^{3-i}_{k}\frac{\partial}{\partial t_{m+k(3-i)}}.
	\end{align*}
\end{enumerate}
\end{example}
	\begin{remark}
		We can deduce relevant cases generated by some known quantum algebras.
		\begin{enumerate}
			\item [(i)] By choosing $\mathcal{R}(u)=(1-q)^{-1}(1-u),$ we obtain:
			the $q$-operators
			\begin{equation}\label{aqa}
			_aW^r_m=-\mathcal{D}^{r-1}_{\mathcal{R}(p^{a},q^{a})}\,x^{m+r-1},\quad m \in\mathbb{Z},\quad r\in\mathbb{Z}_{+},
			\end{equation}
			where  $\mathcal{D}_{q^{a}}$ is the $q$-deformed derivative given by:
			\begin{equation*}
			\mathcal{D}_{q^{a}}f(x)=\frac{f(x)-f(q^a\,x)}{(1-q^a)x},\quad a>1.
			\end{equation*}
			The relation \eqref{aqa} can be rewritten as follows:
			\begin{equation*}
			_aW^r_m=-\, _{a}\mathbb{A}^{r-1}_{m+r-1}\,x^{m}
			=\prod_{i=1}^{r-1}\,T^{q^{a}}_{m+r-i-1},
			\end{equation*}where the $q$-operator $T^{q^{a}}_{m}$ is introduced in \cite{WYLWZ}. Moreover, from the constraints \eqref{qt7}, we derive 
			 the $q$-operator
			\begin{align}\label{qt8}
			_a\bar{W}^r_m&=\sum_{i=0}^{r-1}\mathbb{C}^{i}_{r-1}\,_a\mathbb{A}^{i}_{m+r+\gamma-1}q^{a(r-1-i)(m+r+\gamma-1)} \nonumber\\&\times\sum_{k=1}^{\infty}\frac{(m+k)!}{k!}B_k(t_1,\cdots,t_k)\,_a\mathbb{A}^{r-1-i}_{k}\frac{\partial}{\partial t_{m+k}},
			\end{align}
			with 
			\begin{equation*}
			\mathbb{C}^k_n:= \frac{[n]_q!}{[k]_q![n-k]_q!}, n\geq k,\,\mbox{and}\,\, \mathbb{A}_n^{k}=\left\{\begin{array}{cc}
			[n]_{q}[n-1]_{q}\cdots[n-k+1]_{q},& k\leqslant n,\\
			0,                  &k>n.\end{array}\right.
			\end{equation*}
\item[(ii)] 	Taking $r=2$ in \eqref{qt8}, we obtain the operator given by \cite{WYLWZ}.
			\item[(ii)] For $\mathcal{R}(x,y)=(p-q)^{-1}(x-y),$ we obtain:
			the $(p,q)$-operators
			\begin{eqnarray}\label{apqa}
			_aW^r_m=-\mathcal{D}^{r-1}_{p^{a},q^{a}}\,x^{m+r-1},\quad m \in\mathbb{Z},\quad r\in\mathbb{Z}_{+},
			\end{eqnarray}
			where  $\mathcal{D}_{p^{a},q^{a}}$ is the $(p,q)$-derivative:
			\begin{eqnarray*}
			\mathcal{D}_{p^{a},q^{a}}f(x)=\frac{f(p^a\,x)-f(q^a\,x)}{(p^a-q^a)x},\quad a>1.
			\end{eqnarray*}
			The relation \eqref{apqa} is reduced in the form:
			\begin{eqnarray*}
			_aW^r_m=-\, _{a}\mathbb{A}^{r-1}_{m+r-1}\,x^{m}=\prod_{i=1}^{r-1}\,T^{p^{a},q^{a}}_{m+r-i-1},
			\end{eqnarray*} with $T^{p^{a},q^{a}}_{m}$ the $(p,q)$-operator described in \cite{HMM}. Besides, by using the relation \eqref{qt7}, we deduce the $(p,q)$-operators:
			\begin{align*}
			_a\bar{W}^r_m&=\sum_{i=0}^{r-1}\mathbb{C}^{i}_{r-1}\,_a\mathbb{A}^{i}_{m+r+\gamma-1}q^{a(r-1-i)(m+r+\gamma-1)} \nonumber\\&\times\sum_{k=1}^{\infty}p^{aik}\frac{(m+k)!}{k!}B_k(t_1,\cdots,t_k)\,_a\mathbb{A}^{r-1-i}_{k}\frac{\partial}{\partial t_{m+k}}
			\end{align*}
			where 
			\begin{equation*}
				\mathbb{C}^k_n:= \frac{[n]_{p,q}!}{[k]_{p,q}![n-k]_{p,q}!}, n\geq k,\,\mbox{and}\,\, \mathbb{A}_n^{k}=\left\{\begin{array}{cc}
					[n]_{p,q}[n-1]_{p,q}\cdots[n-k+1]_{p,q},& k\leqslant n,\\
					0,                  &k>n.\end{array}\right.
			\end{equation*}
		\end{enumerate}
	\end{remark}
\section{Concluding remarks}
The generalization of the $W_{1+\infty}$-algebra, the multi-variable $W_{1+\infty}$-algebra and its $n$-algebra from the generalized quantum algebra have been constructed and investigated. Furthermore,  the $\mathcal{R}(p,q)$-$W_{\infty}$-constraints for the  elliptic hermitian  matrix model and a toy model have been derived and presented. Particular cases have been also deduced.
\section*{Acknowledgments}
FM was supported\footnote{``Funded by the Deutsche
	Forschungsgemeinschaft (DFG, German Research Foundation) -- Project ID
	541735537''} by an AIMS-DFG cooperation visit and by a Georg Forster
Fellowship of the Alexander von Humboldt Foundation.
RW was supported\footnote{``Funded by the Deutsche
	Forschungsgemeinschaft (DFG, German Research Foundation) --
	Project ID 427320536 -- SFB 1442, as well as under Germany's
	Excellence Strategy EXC 2044 390685587, Mathematics M\"unster:
	Dynamics -- Geometry -- Structure.''} by the Cluster of Excellence
\emph{Mathematics M\"unster} and the CRC 1442 \emph{Geometry:
	Deformations and Rigidity}.

\end{document}